\documentclass[12pt,letterpaper]{article}%
\usepackage{amsmath}
\usepackage{amsfonts}
\usepackage{amssymb}
\usepackage{nopageno}
\usepackage{graphicx}%
\newtheorem{theorem}{Theorem}

\newtheorem{claim}[theorem]{Claim}

\newtheorem{remark}[theorem]{Remark}

\makeatletter
\def\@oddhead{
\vbox{
\hbox to\hsize{\oddmarkA \oddmarkB \hfill \oddmarkC}
}
}
\makeatother
\def\oddmarkA{{\bf }{}}
\def\oddmarkB{}
\def\oddmarkC{\thepage}
\begin{document}

\begin{center}
\bigskip{\LARGE Turbulence and the Navier-Stokes Equations}

\textbf{R. M. Kiehn}

\textit{Emeritus Professor of Physics, University of Houston}

\textit{\ Retired to Mazan, France}

\textit{\ http://www.cartan.pair.com}
\end{center}

\bigskip

\begin{quote}
\textbf{Abstract: \ }The concept of continuous topological evolution, based
upon Cartan's methods of exterior differential systems, is used to develop a
topological theory of non-equilibrium thermodynamics, within which there exist
processes that exhibit continuous topological change and thermodynamic
irreversibility. \ The technique furnishes a universal, topological foundation
for the partial differential equations of hydrodynamics and electrodynamics;
\ the technique does not depend upon a metric, connection or a variational
principle. \ Certain topological classes of solutions to the Navier-Stokes
equations are shown to be equivalent to thermodynamically irreversible processes.
\end{quote}

\bigskip

{\Large Prologue}

\begin{center}
\textbf{THE POINT OF DEPARTURE \ }
\end{center}

\subparagraph{ \ }

This presentation summarizes a portion of 40 years of research
interests\footnote{This work is summarized in a series of reference monographs
\cite{vol1}, \cite{vol2}, \cite{vol3}, \cite{vol4}, \cite{vol5} which have
been constructed and updated from numerous publications. These volumes contain
many examples and proofs of the basic concepts. \ } in applied physics from a
perspective of continuous topological evolution. \ The motivation for the past
and present effort continues to be based on the recognition that topological
evolution (not geometrical evolution) is required if non-equilibrium
thermodynamic systems and irreversible turbulent processes are to be
understood without the use of statistics. \ This essay is written (by an
applied physicist) as an alternative response to the (more mathematical)
challenge of the Clay Institute regarding the properties of the Navier-Stokes
equations and their relationship to hydrodynamic turbulence. \ To replicate a
statement made by the Clay Institute:

\begin{quote}
\ "The challenge is to make substantial progress toward a mathematical theory
which will unlock the secrets hidden in the Navier-Stokes equations." \ \ 
\end{quote}

The point of departure starts with a topological (not
statistical)\ formulation of Thermodynamics, which furnishes a universal
foundation for the Partial Differential Equations of classical hydrodynamics
and electrodynamics \cite{vol1}. \ The topology that is of significance is
defined in terms of Cartan's topological structure
\cite{rmkarXiv:math-ph/0101033}, which can be constructed from an exterior
differential 1-form, $A$, defined on a pre-geometric domain of base variables.
\ The topological method extends the classical geometrical approach to the
study of non-equilibrium thermodynamic systems. \ 

\begin{claim}
The topological method permits the conclusion that among the solutions to the
Navier-Stokes equations there are C2 smooth, thermodynamically irreversible
processes which permit description of topological change and the decay of turbulence.
\end{claim}

In addition, the method permits examples to be constructed showing the
difference between certain piecewise-linear processes which are reversible,
but which are different from certain smooth processes which are irreversible.
\ Such concepts (of smoothness) seem to be of direct interest to the challenge
of the Clay Institute, and are to be associated with the fact that there are
differences between piecewise linear, smooth, and topological manifolds\ (see
p. 106, \cite{Stewart}).

\ However, the methods of topological thermodynamics go well beyond these
types of questions. \ In particular, the methods permit non-statistical
engineering design criteria to be developed for non-equilibrium systems. \ The
theory of Topological Thermodynamics, based upon Continuous Topological
Evolution \cite{rmkarXiv:math-ph/0101032} of Cartan's topological structure,
can explain why topologically coherent, compact structures, far from
equilibrium, will emerge as long-lived artifacts of thermodynamically
irreversible, turbulent, continuous processes. \ I want to present the idea that:

\begin{theorem}
The Pfaff Topological Dimension (PTD) of a Thermodynamic System can change
dynamically and continuously via irreversible dissipative processes from a
non-equilibrium turbulent state of PTD = 4 to an excited \textquotedblleft
topologically stationary, but excited,\textquotedblright\ state of PTD = 3,
which is still far from equilibrium! \ The PTD=3 state admits an extremal
Hamiltonian evolutionary process which, if dominant, produces a relatively
long lifetime.
\end{theorem}

\ There exist C2 smooth processes that can describe the topological evolution
from an Open non-equilibrium turbulent domain of Pfaff Topological Dimension 4
to Closed, but non-equilibrium, domains of Pfaff Topological Dimension 3, and
ultimately to equilibrium domains of Pfaff dimension 2 or less. \ The
Topological domains of Pfaff Topological Dimension 3 emerge via
thermodynamically irreversible, dissipative processes as topologically
coherent, deformable defects, embedded in the turbulent environment of Pfaff
Topological Dimension 4.

Now I am well aware of the fact that Thermodynamics (much less Topological
Thermodynamics) is a topic often treated with apprehension. \ In addition, I
must confess, that as undergraduates at MIT we used to call the required
physics course in Thermodynamics, The Hour of Mystery! \ Let me present a few
quotations (taken from Uffink, \cite{Uffink}) that describe the apprehensive
views of several very famous scientists:

\begin{quote}
Any mathematician knows it is impossible to understand an elementary course in
thermodynamics ....... V. Arnold 1990.

\bigskip

It is always emphasized that thermodynamics is concerned with reversible
processes and equilibrium states, and that it can have nothing to do with
irreversible processes or systems out of equilibrium ......Bridgman 1941

\bigskip

No one knows what entropy really is, so in a debate (if you use the term
entropy) you will always have an advantage ...... Von Neumann (1971)\bigskip
\end{quote}

On the other hand Uffink states:

\begin{quote}
Einstein, ..., remained convinced throughout his life that thermodynamics is
the only universal physical theory that will never be overthrown.
\end{quote}

I wish to demonstrate that from the point of view of Continuous Topological
Evolution (which is based upon Cartan's theory of exterior differential forms)
many of the mysteries of non-equilibrium thermodynamics, irreversible
processes, and turbulent flows, can be resolved. \ In addition, the
non-equilibrium methods can lead to many new processes and patentable devices
and concepts.

There are many intuitive, yet disputed, definitions of what is meant by
turbulence, but the one property of turbulence that everyone agrees upon is
that turbulent evolution in a fluid is a thermodynamic irreversible process.
\ Isolated or equilibrium thermodynamics can be defined on a 4D space-time
variety in terms of a connected Cartan topology of Pfaff Topological Dimension
2 or less. \ Non-equilibrium thermodynamics can be constructed in terms of
disconnected Cartan topology of Pfaff Topological Dimension of 3 or more. \ As
irreversibility requires a change in topology, the point of departure for this
article will be to use the thermodynamic theory of continuous topological
evolution in 4D space-time. \ It will be demonstrated, by example, that the
non-equilibrium component of the Cartan topology can support topological
change, thermodynamic irreversible processes and turbulent solutions to the
Navier-Stokes equations, while the equilibrium topological component cannot.
\ In addition, it will be demonstrated that complex isotropic
\textit{macroscopic} Spinors are the source of topological fluctuations and
irreversible processes in the topological dynamics of non-equilibrium systems.
\ This, perhaps surprising, fact has been ignored by almost all researchers in
classical hydrodynamics who use classic real vector analysis\ and symmetries
to produce conservation laws, which do not require Spinor components. \ The
flaw in such symmetrical based theories is that they describe evolutionary
processes that are time reversible. \ Time irreversibility requires
topological change.

\begin{center}
\ \textbf{EXTERIOR DIFFERENTIAL FORMS }

overcomes the

\textbf{LIMITATIONS of REAL VECTOR\ ANALYSIS }
\end{center}

During the period 1965-1992 it became apparent that new theoretical
foundations were needed to describe non-equilibrium systems and continuous
irreversible processes - which require topological (not geometrical)
evolution. \ I selected Cartan's methods of exterior differential topology to
encode Continuous Topological Evolution. \ The reason for this choice is that
many years of teaching experience indicated that such methods were rapidly
learned by all research scientists and engineers. \ In short:

\begin{enumerate}
\item Vector and Tensor analysis is not adequate to study the evolution of
topology. \ The tensor constraint of diffeomorphic equivalences implies that
the topology of the initial state must be equal to the topology of the final
state. \ Turbulence is a thermodynamic, irreversible process which can not be
described by tensor fields alone.

\item However, Cartan's methods of exterior differential systems and the
topological perspective of Continuous Topological Evolution (not geometrical
evolution) CAN be used to construct a theory of non-equilibrium thermodynamic
systems and irreversible processes.

\item Bottom Line: \ Exterior differential forms carry topological information
and can be used to describe topological change induced by processes. \ Real
"Vector" directionfields alone cannot describe processes that cause
topological change; \ but Spinor directionfields can.
\end{enumerate}

A cornerstone of classic Vector (tensor)\ analysis is the constraint of
functional equivalence with respect to diffeomorphisms. \ However,
diffeomorphisms are a subset of a homeomorphisms, and homeomorphisms preserve
topology. \ Hence to study topological change, Vector (tensor) analysis is not
adequate. \ In topological thermodynamics, processes are defined in terms of
directionfields which may or may not be tensors. \ The ubiquitous concepts of
1-1 diffeomorphic equivalence, and non-zero congruences, for the eigen
directionfields of symmetric matrices do not apply to the eigen
directionfields of antisymmetric matrices. \ The eigen direction fields of
antisymmetric matrices (which are equivalent to Cartan's isotropic Spinors)
may be used to define components of a thermodynamic process, but such Spinors
have a null congruence (zero valued quadratic form), admit chirality, and are
not 1-1. \ Where classic geometric evolution is described in terms of
symmetries and conservation laws, topological evolution is described in terms
of antisymmetries. \ 

Cartan's theory of exterior differential forms is built over completely
antisymmetric structures, and therefore is the method of choice for studying
topological evolution. \ The exterior differential defines limit sets; the Lie
differential defines continuous topological evolution. \ The concept of
Spinors arise naturally in theories using Cartan's methods of exterior
differential forms; \ i.e., Spinors are not added to the theory ad hoc. \ The
Cartan theory of extended differential forms can be used to study topological
change. \ The word \textit{extended} is used to emphasize the fact that
differential forms are functionally well defined with respect a larger class
of transformations\ than those used to define tensors. \ Extended differential
forms behave as scalars with respect to C1 maps which do not have an inverse,
much less an inverse Jacobian. \ Both the inverse map and the inverse Jacobian
are required by a diffeomorphism.\ \ The exterior differential form on the
final state of such C1 non-invertible maps permits the functional form of the
differential form on the initial state to be functionally well defined in a
retrodictive, pullback sense - not just at a point, but over a neighborhood. \ 

\begin{theorem}
Tensor fields can be neither retrodicted nor predicted in functional form by
maps that are not diffeomorphisms \cite{rmkretro}.
\end{theorem}

\begin{center}
\bigskip

\textbf{CONTINUOUS\ TOPOLOGICAL\ EVOLUTION}
\end{center}

\subparagraph{Objectives of CTE}

The objectives of the theory of Continuous Topological Evolution are to:

\begin{enumerate}
\item Establish the long sought for connection between irreversible
thermodynamic processes and dynamical systems -- without statistics! \ 

\item Demonstrate the connection between thermodynamic irreversibility and
Pfaff Topological Dimension equal to 4. \ The result suggests that
\textquotedblleft2-D Turbulence is a myth\textquotedblright\ for it is a
thermodynamic system of Pfaff Topological Dimension equal to 3
\cite{rmk2dmyth}.

\item Demonstrate that topological thermodynamics leads to universal
topological equivalences between Electromagnetism, Hydrodynamics, Cosmology,
and Topological Quantum Mechanics.

\item Demonstrate that Cartan's methods of exterior differential forms permits
important topological concepts to be displayed in a useful, engineering format.
\end{enumerate}

\subparagraph{New Concepts deduced from CTE}

The theory of Continuous Topological Evolution introduces several new
important concepts that are not apparent in a geometric equilibrium analysis.

\begin{enumerate}
\item Continuous Topological Evolution is the dynamical equivalent of the
FIRST LAW OF THERMODYNAMICS.

\item The Pfaff Topological Dimension, PTD, is a topological property
associated with any Cartan exterior differential 1-form, $A$. \ The PTD can
change via topologically continuous processes.

\item Topological Torsion is a 3-form (on any 4D geometrical domain) that can
be used to describe irreversible processes. \ As a 4D non-equilibrium
direction field it is completely determined by the coefficient functions that
encode the thermodynamic system. \ Other process directionfields are
determined by the system topology based upon the 1-form of Action, $A$, and
the refinement based on the topology of the 1-form of work, $W$.

\item Closed thermodynamic topological defects of Pfaff Topological Dimension
3 can emerge from Open thermodynamic systems of Pfaff Topological Dimension 4
by means of irreversible dissipative processes that represent topological
evolution and change. \ When the topologically coherent defect structures
emerge, their evolution can be dominated by a Hamiltonian component (modulo
topological fluctuations), which maintains the topological deformation
invariance, and yields hydrodynamic wakes \cite{rmkpoitier} and other Soliton
structures. \ These objects are of Pfaff Topological Dimension 3 and are far
from equilibrium. \ They behave as if they were "stationary excited" states
above the equilibrium ground state. \ Falaco Solitons are an easily reproduced
hydrodynamic example that came to my attention in 1986 \cite{vol2}
\cite{rmkarXiv:gr-qc/0101098} .
\end{enumerate}

\begin{center}
\textbf{PRESENTATION\ OUTLINE}
\end{center}

The essay is constructed in several sections:

\paragraph{Section 1. Topological Thermodynamics}

In Section 1, the concepts of topological thermodynamics in a space-time
variety are reviewed (briefly) in terms of Cartan's method of exterior
differential forms. \ A thermodynamic system is encoded in terms of a 1-form
of Action, $A$. \ Thermodynamic processes are encoded in terms of the Lie
differential with respect to a directionfield, $V$, acting on the 1-form, $A$,
to produce a 1-form, $Q$. \ The process directionfield can have Vector and
Spinor components. \ The definition of the Lie differential is a statement of
cohomology and defines $Q$ as the composite of a 1-form, $W$, and a perfect
differential, $dU$. \ The formula abstractly represents a dynamical version of
the First Law of Thermodynamics. \ 

The existence of a 1-form on a 4D space-time variety generates a Cartan
topology. \ If the Pfaff Topological (not geometrical) Dimension of the 1-form
of Action, $A$, is 2 or less, then the thermodynamic system is an isolated or
equilibrium system on the 4D variety. \ If the Pfaff Topological Dimension of
$A$ is greater than 3, then the system is a non-equilibrium system on the 4D
variety. \ Examples of systems of Pfaff Topological Dimension 4 which admit
processes which are thermodynamically irreversible are given in the reference
monographs (see footnote page 1).

\paragraph{Section 2.\ Applications}

In Section 2, the abstract formalism will be given a specific realization
appropriate for fluids in general. \ First, an electromagnetic format will be
described because my teaching experience has demonstrated that the concepts of
non-equilibrium phenomena are more readily recognized in an electromagnetic
format. \ Then it will be demonstrated how the PDE's representing the
Hamiltonian version of the hydrodynamic Lagrange-Euler equations arise from
the constraint that the work 1-form, $W$, should vanish (Pfaff Topological
Dimension of $W=0$). \ The Bernoulli flow will be obtained by constraining the
thermodynamic Work 1-form to be exact, $W=d\Theta$ (Pfaff Topological
Dimension 1), and the Helmholtz flow will follow from the constraint that the
thermodynamic Work 1-form be closed, but not necessarily exact, $dW=0$. \ Such
reversible dynamical processes belong to the connected component of the Work
1-form; irreversible processes belong to the disconnected component of the
Work 1-form. \ 

\paragraph{Section 3. The Navier-Stokes system}

In Section 3, the topological constraints of isolated equilibrium systems will
be relaxed to produce more general PDE's defining the topological evolution of
the system relative to an applied process. \ These relaxed topological
constraints will include the partial differential equations known as the
Navier-Stokes equations. \ The method used will be to augment the topology
induced by the 1-form of Action, $A$, by studying the topological refinements
induced by the 1-form of Work, $W$. \ It will be demonstrated that when the
Pfaff Topological Dimension of $A$ and $W$ and $Q$ are $4$, there exist C2
solutions (processes) to the Navier-Stokes equations which are
thermodynamically irreversible (the most significant property of turbulent
flow). \ An interesting result is the set of conditions on solutions of the
Navier-Stokes equations that produce an adiabatic \textit{irreversible} flow. \ 

Those topological refinements of the Work 1-form, required to include the
Navier-Stokes equations, can be related directly to the concept of macroscopic
Spinors. \ Macroscopic, complex Spinor solutions occur naturally in terms of
the eigendirection fields of (real) antisymmetric matrices with non-zero
eigenvalues, whenever the thermodynamic Work 1-form is not zero. \ Spinors can
also be associated with topological fluctuations of position and velocity
about kinematic perfection generated by 1-parameter groups. \ These
topological fluctuations are presumed to be representations of pressure and temperature.

\paragraph{Section 4. \ Closed States of Topological Coherence embedded as
deformable defects in Turbulent Domains}

One of the key interests of the Clay problem has to do with the smoothness of
the solutions to the Navier-Stokes equations. \ In Section 4, the problem will
be attacked from the point of view of thermodynamics. \ First, the properties
of the different species of topological defects will be discussed. \ These
defects are non-equilibrium closed domains (of $PTD=3$) which can emerge by C2
smooth irreversible process in open domains (of $PTD=4 $), as excited states
far from equilibrium, yet with long relative lifetimes. \ Falaco Solitons are
an easily reproduced \textit{experimental} example of such topological
defects, and are discussed in detail in \cite{vol2}. \ 

The properties of two different species of $PTD=3$ defect domains will be
given in detail. \ In addition, an analytic solution of a thermodynamically
irreversible process that causes the defect domain to emerge will be displayed.

Finally, an example will be given where by combinations of Spinor solutions
produce piecewise linear processes. \ These piecewise linear processes are
thermodynamically reversible, while the Spinor solutions of which they are
composed are not.

\paragraph{\ Section 5. \ Topological Fluctuations and Spinors}

In Section 5, a few concluding remarks will be made about the ongoing research
concerning topological fluctuations, as generated by Spinors. \ The methods of
fiber bundle theory are used extend the 4D thermodynamic domain. \ Such
topological fluctuations can be associated with fluid pressure and
temperature. \ \newpage

\section{Topological Thermodynamics}

\subsection{The Axioms of Topological Thermodynamics}

The topological methods used herein are based upon Cartan's theory of exterior
differential forms. \ The thermodynamic view assumes that the physical systems
to be studied can be encoded in terms of a 1-form of Action Potentials (per
unit source, or, per mole), $A,$ on a four-dimensional variety of ordered
independent variables, $\{\xi^{1},\xi^{2},\xi^{3},\xi^{4}\}.$ \ The variety
supports a differential volume element $\Omega_{4}=d\xi^{1}\symbol{94}d\xi
^{2}\symbol{94}d\xi^{3}\symbol{94}d\xi^{4}.$ \ This statement implies that the
differentials of the $\mu=4$ base variables are functionally independent. \ No
metric, no connection, no constraint of gauge symmetry is imposed upon the
four-dimensional pre-geometric variety. \ \ Topological constraints can be
expressed in terms of exterior differential systems placed upon this set of
base variables $\cite{Bryant}.$ \ 

In order to make the equations more suggestive to the reader, the symbolism
for the variety of independent variables will be changed to the format
$\{x,y,z,t\},$ but be aware that no constraints of metric or connection are
imposed upon this variety, at this, thermodynamic, level. \ For instance, it
is NOT assumed that the variety is spatially Euclidean. \ 

With this notation, the Axioms of Topological Thermodynamics can be summarized as:

\begin{quote}
\textbf{Axiom 1}. \ \textit{Thermodynamic physical systems can be encoded in
terms of a 1-form of covariant Action Potentials, }$A_{\mu}(x,y,z,t...),$%
\textit{\ on a four-dimensional abstract variety of ordered independent
variables, }$\{x,y,z,t\}$. \ \textit{The variety supports differential volume
element} $\Omega_{4}=dx\symbol{94}dy\symbol{94}dz\symbol{94}dt.$

\textbf{Axiom 2.} \ \textit{Thermodynamic processes are assumed to be encoded,
to within a factor, }$\rho(x,y,z,t...)$\textit{, in terms of a contravariant
Vector and/or complex Spinor directionfields, symbolized as }$V_{4}(x,y,z,t)$.

\textbf{Axiom 3. }\ \textit{Continuous Topological Evolution of the
thermodynamic system can be encoded in terms of Cartan's magic formula (see p.
122 in \cite{Marsden}). \ The Lie differential with respect to the process,
}$\rho V_{4}$\textit{, when applied to an exterior differential 1-form of
Action, \thinspace}$A=A_{\mu}dx^{\mu}$\textit{, is equivalent, abstractly, to
the first law of thermodynamics.}%
\begin{align}
\text{ Cartan's Magic Formula }L_{(\rho\mathbf{V}_{4})}A  & =i(\rho
\mathbf{V}_{4})dA+d(i(\rho\mathbf{V}_{4})A),\\
\text{First Law }  & :W+dU=Q,\\
\text{Inexact Heat 1-form\ \ }Q  & =W+dU=L_{(\rho\mathbf{V}_{4})}A,\\
\text{Inexact Work 1-form\ }W  & =i(\rho\mathbf{V}_{4})dA,\\
\text{Internal Energy \ }U  & =i(\rho\mathbf{V}_{4})A.
\end{align}

\textbf{Axiom 4.} \ \textit{Equivalence classes of systems and continuous
processes can be defined in terms of the Pfaff Topological Dimension and
topological structure generated by of the 1-forms of Action, }$A$, Work, $W$,
and Heat, $Q$.

\textbf{Axiom 5. }If $Q\symbol{94}dQ\neq0$, \textit{then the thermodynamic
process is irreversible.}
\end{quote}

\subsection{Cartan's Magic Formula $\approx$ First Law of Thermodynamics}

The Lie differential (not Lie derivative) is the fundamental generator of
Continuous Topological Evolution. \ When acting on an exterior differential
1-form of Action, \thinspace$A=A_{\mu}dx^{\mu}$, Cartan's magic (algebraic)
formula is equivalent \textit{abstractly} to the first law of thermodynamics:
\ \
\begin{align}
L_{(\rho\mathbf{V}_{4})}A  & =i(\rho\mathbf{V}_{4})dA+d(i(\rho\mathbf{V}%
_{4})A),\\
& =W+dU=Q.
\end{align}
In effect, Cartan's magic formula leads to a topological basis of
thermodynamics, where the thermodynamic Work, $W$, thermodynamic Heat, $Q$,
and the thermodynamic internal energy, $U$, are defined\textit{\ dynamically}
in terms of Continuous Topological Evolution. \ \ In effect, the First Law is
a statement of Continuous Topological Evolution in terms of deRham cohomology
theory; \ the difference between two non-exact differential forms is equal to
an exact differential, $Q-W=dU$. \ 

My recognition (some 30 years ago) of this correspondence between the Lie
\textit{differential} and the First Law of thermodynamics has been the corner
stone of my research efforts in applied topology.

It is important to realize that the Cartan formula is to be interpreted
algebraically. \ Many textbook presentations of the Cartan-Lie differential
formula presume a dynamic constraint, such that the vector field
$\mathbf{V}_{4}(x,y,z,t)\,$\ be the generator of a single parameter group.
\ If true, then the topological constraint of Kinematic Perfection cn be
established as an exterior differential system of the format:%
\begin{equation}
\text{\textbf{Kinematic Perfection :\ \ }}\mathbf{\ \ }dx^{k}-\mathbf{V}%
^{k}dt\Rightarrow0.
\end{equation}
The topological constraint of Kinematic Perfection, in effect, defines (or
presumes) a limit process. \ This constraint leads to the concept of the Lie
\textit{derivative\footnote{Professor Zbigniew Oziewicz told me that
Slebodzinsky was the first to formulate the idea of the Lie derivative in his
thesis (in Polish).}} of the 1-form $A.$ \ The evolution then is represented
by the infinitesimal propagation of the 1-form, $A$, down the flow lines
generated by the 1-parameter group. \ Cartan called this set of flow lines
"the tube of trajectories". \ 

However, such a topological, kinematic constraint is \textit{not} imposed in
the presentation found in this essay; \ the directionfield, $\mathbf{V}_{4}$,
may have multiple parameters. \ This observation leads to the important
concept of topological fluctuations (about Kinematic Perfection), such as
given by the expressions: \
\begin{align}
\text{\textbf{Topological }}  & \mathbf{:}\text{{}}\mathbf{~\ }%
\text{\textbf{Fluctuations\ }}\mathbf{\ \ }\nonumber\\
(dx^{k}-\mathbf{V}^{k}dt)  & =\Delta\mathbf{x}^{k}\neq0,\text{ \ \ \ \ \ (}%
\thicksim\text{Pressure)}\\
(dV^{k}-\mathbf{A}^{k}dt)  & =(\Delta\mathbf{V}^{k})\neq0,\text{ \ (}%
\thicksim\text{Temperature)}\\
d(\Delta\mathbf{x}^{k})  & =-(d\mathbf{V}^{k}-\mathbf{A}^{k}dt)\symbol{94}%
dt=-(\Delta\mathbf{V}^{k})\symbol{94}dt,
\end{align}
In this context it is interesting to note that in Felix Klein's discussions
\cite{KleinF}\ of the development of calculus, he says

\begin{quote}
"The primary thing for him (Leibniz) was not the differential quotient (the
derivative) thought of as a limit. \ The differential, dx, of the variable x
had for him (Leibniz) actual existence..."\ \ \ \ 
\end{quote}

The Leibniz concept is followed throughout this presentation. It is important
for the reader to remember that the concept of a differential form is
different from the concept of a derivative, where a (topological) limit has
been defined, thereby constraining the topological evolution. \ \ \ \ 

The topological methods to be described below go beyond the notion of
processes which are confined to equilibrium systems of kinematic perfection.
\ Non-equilibrium systems and processes which are thermodynamically
irreversible, as well as many other classical thermodynamic ideas, can be
formulated in precise mathematical terms using the topological structure and
refinements generated by the three thermodynamic 1-forms, $A,\ W,\ $and $Q$. \ 

\subsection{The Pfaff Sequence and the Pfaff Topological Dimension}

\subsubsection{The Pfaff Topological Dimension of the System 1-form, $A$}

It is important to realize that the Pfaff Topological Dimension of the system
1-form of Action, $A$, determines whether the thermodynamic system is Open,
Closed, Isolated or Equilibrium. \ Also, it is important to realize that the
Pfaff Topological Dimension of the thermodynamic Work 1-form, $W$, determines
a specific category of reversible and/or irreversible processes. \ It is
therefore of some importance to understand the meaning of the Pfaff
Topological Dimension of a 1-form. \ Given the functional format of a general
1-form, $A$, on a 4D variety it is an easy step to compute the Pfaff Sequence,
using one exterior differential operation, and several algebraic exterior
products. \ For a differential 1-form, $A$, defined on a geometric domain of 4
base variables, the Pfaff Sequence is defined as:
\begin{equation}
\text{\textbf{Pfaff Sequence} \ }\{A,dA,A\symbol{94}dA,dA\symbol{94}%
dA...\}\label{PS}%
\end{equation}
It is possible that over some domains, as the elements of the sequence are
computed, one of the elements (and subsequent elements) of the Pfaff Sequence
will vanish. \ The number of non-zero elements in the Pfaff Sequence
(PS)\ defines the Pfaff Topological Dimension (PTD) of the specified
1-form\footnote{The Pfaff Topological dimension has been called the "class" of
a 1-form in the old literature. \ I prefer the more suggestive name.}.
\ Modulo singularities, the Pfaff Topological Dimension determines the minimum
number $M$ of $N$ functions\ of base variables ($N\geq M$) required to define
the topological properties of the connected component of the 1-form $A$. \ 

The Pfaff Topological Dimension of the 1-form of Action, $A$, can be put into
correspondence with the four classic topological structures of thermodynamics.
Equilibrium, Isolated, Closed, and Open systems. \ The classic thermodynamic
interpretation is that the first two structures do not exchange mass (mole
numbers) or radiation with their environment. \ The Closed structure can
exchange radiation with its environment but not mass (mole numbers). \ The
Open structure can exchange both mass and radiation with its environment.
\ The following table summarizes these properties. \ For reference purposes, I
have given the various elements of the Pfaff sequence specific names: \ 

\begin{center}
\bigskip$%
\begin{array}
[c]{ccccc}%
\begin{array}
[c]{c}%
\text{Topological}\\
\text{p-form name}%
\end{array}
&
\begin{array}
[c]{c}%
\text{\textbf{PS} }\\
\text{element}%
\end{array}
& \text{\textbf{Nulls}} & \text{\textbf{PTD}} &
\begin{array}
[c]{c}%
\text{Thermodynamic}\\
\text{system}%
\end{array}
\\
\text{\textbf{Action}} & A & ~dA=0 & 1 & \text{\textbf{Equilibrium}}\\
\text{\textbf{Vorticity}} & dA~ & A\symbol{94}dA=0 & 2 &
\text{\textbf{Isolated}}\\
\text{\textbf{Torsion}} & A\symbol{94}dA & dA\symbol{94}dA=0 & 3 &
\text{\textbf{Closed}}\\
\text{\textbf{Parity}} & dA\symbol{94}dA & - & 4 & \text{\textbf{Open}}%
\end{array}
$

\textbf{Table 1 \ Applications of the Pfaff Topological Dimension.}
\end{center}

The four thermodynamic systems can be placed into two disconnected topological
categories. \ If the Pfaff Topological Dimension of $A$ is 2 or less, the
first category is determined by the closure (or differential ideal) of the
1-form of Action, $A\cup dA$. $\ $This Cartan topology is a connected
topology. \ In the case that the Pfaff Topological Dimension is greater than
2, the Cartan topology is based on the union of two closures, $\{A\cup
dA~\cup~A\symbol{94}dA\cup dA\symbol{94}dA\}$, and is a disconnected topology. \ 

It is a topological fact that there exists a (topologically) continuous C2
process from a disconnected topology to a connected topology, but there does
not exist a C2 continuous process from a connected topology to a disconnected
topology. \ This fact implies that topological change can occur continuously
by a "pasting" processes representing the \textit{decay}\ of turbulence by
"condensations" from non-equilibrium to equilibrium systems. \ On the other
hand, the \textit{creation} of Turbulence involves a discontinuous (non C2)
process of "cutting" into parts. \ This warning was given long ago
\cite{RMKsectam} to prove that computer analyses that smoothly match value and
slope will not replicate the \textit{creation} of turbulence, but can
faithfully replicate the \textit{decay} of turbulence.

\subsubsection{The Pfaff Topological Dimension of the Thermodynamic Work
1-form, $W$}

The topological structure of the thermodynamic Work 1-form, $W$, can be used
to refine the topology of the physical system; \ recall that the physical
system is encoded by the Action 1-form, $A$. \ 

\begin{claim}
The PDE's that represent the system dynamics are determined by the Pfaff
Topological Dimension of the 1-form of Work, $W$, and the 1-form of Action,
$A$, that encodes the physical system.\ 
\end{claim}

The Pfaff Topological Dimension of the thermodynamic Work 1-form depends upon
both the physical system, $A$, and the process, $\mathbf{V}_{4}$. \ In
particular if the Pfaff Dimension of the thermodynamic Work 1-form is zero,
($W=0$), then system dynamics is generated by an extremal vector field which
admits a Hamiltonian realization. \ However, such extremal direction fields
can occur only when the Pfaff Topological Dimension of the system encoded by
$A$ is odd, and equal or less than the geometric dimension of the base
variables. \ 

For example, if the geometric dimension is 3, and the Pfaff Topological
Dimension of $A$ is 3, then there exists a unique extremal field on the
Contact manifold defined by $dA$. \ This unique directionfield is the unique
eigen directionfield of the 3x3 antisymmetric matrix (created by the 2-form
$F=dA)$ with eigenvalue equal to zero. \ 

If the geometric dimension is 4, and the Pfaff Topological Dimension of $A$ is
3, then there exists a two extremal fields on the geometric manifold. \ These
directionfields are those generated as the eigen directionfields of the 4x4
antisymmetric matrix (created by the 2-form $F=dA)$ with eigenvalue equal to
zero. \ 

If the geometric dimension is 4, and the Pfaff Topological Dimension of $A$ is
4, then there do not exist extremal fields on the Symplectic manifold defined
by $dA$. \ All of the eigen directionfields of the 4x4 antisymmetric matrix
(created by the 2-form $F=dA)$ are complex isotropic spinors with pure
imaginary eigenvalues not equal to zero. \ 

\subsection{Topological Torsion and other Continuous Processes.}

\subsubsection{Reversible Processes}

Physical Processes are determined by directionfields\footnote{Which include
both vector and spinor fields.} with the symbol, $\mathbf{V}_{4}$, to within
an arbitrary function, $\rho$. \ There are several classes of direction fields
that are defined as follows \cite{Klein}:%

\begin{align}
\text{\textbf{Associated Class}}  & \mathbf{:}\text{}i(\rho\mathbf{V}%
_{4})A=0,\\
\text{\textbf{Extremal Class}}  & \mathbf{:}\text{}i(\rho\mathbf{V}%
_{4})dA=0,\\
\text{\textbf{Characteristic Class}}  & \mathbf{:}\text{}i(\rho\mathbf{V}%
_{4})A=0,\\
\text{and }  & \text{: \ }i(\rho\mathbf{V}_{4})dA=0,\\
\text{\textbf{Helmholtz Class}}  & \mathbf{: \ \ }\text{}d(i(\rho
\mathbf{V}_{4})dA)=0,
\end{align}

Extremal Vectors (relative to the 1-form of Action, $A$) produce zero
thermodynamic work, $W=i(\rho\mathbf{V}_{4})dA=0$, and admit a Hamiltonian
representation. \ Associated Vectors (relative to the 1-form of Action, $A$)
can be adiabatic if the process remains orthogonal to the 1-form, $A.$
\ Helmholtz processes (which include Hamiltonian processes, Bernoulli
processes and Stokes flow) conserve the 2-form of Topological vorticity, $dA.
$ \ All such processes are thermodynamically reversible. \ Many examples of
these systems are detailed in the reference monographs (see footnote on page 1).

\subsubsection{Irreversible Processes}

There is one directionfield that is uniquely defined by the coefficient
functions of the 1-form, $A$, that encodes the thermodynamic system on a
4D\ geometric variety. \ This vector exists only in non-equilibrium systems,
for which the Pfaff Topological Dimension of $A$ is 3 or 4. \ This 4 vector is
defined herein as the topological Torsion vector, $\mathbf{T}_{4}$.\ \ To
within a factor, this directionfield\footnote{A direction field is defined by
the components of a vector field which establish the "line of action" of the
vector in a projective sense. \ An arbitrary factor times the direction field
defines the same projective line of action, just reparameterized. \ In metric
based situations, the arbitrary factor can be interpreted as a renormalization
factor.} has the four coefficients of the 3-form $A\symbol{94}dA,$ with the
following properties:%

\begin{align}
\text{\textbf{Properties of }}  & \mathbf{:}\text{\textbf{Topological Torsion}
}\mathbf{T}_{4}\text{ on }\Omega_{4}\label{PROPT4}\\
i(\mathbf{T}_{4})\Omega_{4}  & =i(\mathbf{T}_{4})dx\symbol{94}dy\symbol{94}%
dz\symbol{94}dt=A\symbol{94}dA,\\
W  & =i(\mathbf{T}_{4})dA=\sigma\ A,\\
dW  & =d\sigma\symbol{94}A+\sigma dA=dQ\\
U  & =i(\mathbf{T}_{4})A=0,\text{ \ }\mathbf{T}_{4}\text{ is associative}\\
i(\mathbf{T}_{4})dU  & =0\text{ \ \ }\\
i(\mathbf{T}_{4})Q  & =0\text{ \ \ \ \ \ \ \ }\mathbf{T}_{4}\text{ is
adiabatic}\\
L_{(\mathbf{T}_{4})}A  & =\sigma\ A,\text{ \ \ }\mathbf{T}_{4}\text{ is
homogeneous }\\
L_{(\mathbf{T}_{4})}dA  & =d\sigma\symbol{94}A+\sigma dA=dQ,\\
Q\symbol{94}dQ  & =L_{(\mathbf{T}_{4})}A\symbol{94}L_{(\mathbf{T}_{4}%
)}dA=\sigma^{2}A\symbol{94}dA\neq0,\text{ \ }\label{testqdq2}\\
dA\symbol{94}dA  & =d(A\symbol{94}dA)=d\{(i(\mathbf{T}_{4})\Omega
_{4}\}=(div_{4}\mathbf{T}_{4})\Omega_{4},\\
L_{(\mathbf{T}_{4})}\Omega_{4}  & =d\{(i(\mathbf{T}_{4})\Omega_{4}%
\}=(2\sigma)\Omega_{4},
\end{align}

If the Pfaff Topological Dimension of $A$ is 4 (an Open thermodynamic system),
then $\mathbf{T}_{4}$ has a non-zero 4 divergence, $(2\sigma)$, representing
an expansion or a contraction of the 4D\ volume element $\Omega_{4}$.\ \ \ The
Heat 1-form, $Q$, generated by the process, $\mathbf{T}_{4}$, is NOT
integrable. \ $Q$ is of Pfaff Topological Dimension greater that 2, when
$\sigma\neq0.$ \ Furthermore the $\mathbf{T}_{4}$ process is locally adiabatic
as the change of internal energy in the direction of the process path is zero.
\ Therefore, in the Pfaff Topological Dimension 4 case, where $dA\symbol{94}%
dA\neq0$, the $\mathbf{T}_{4}$ direction field represents an
\textit{irreversible, adiabatic process}. \ 

When $\sigma$ is zero \textit{and} $d\sigma=0$, but $A\symbol{94}dA\neq0,$ the
Pfaff Topological Dimension of the system is 3 (a Closed thermodynamic
system). \ In this case, the $\mathbf{T}_{4}$ direction field becomes a
characteristic vector field which is both extremal and associative, and
induces a Hamilton-Jacobi representation (the ground state of the system for
which $dQ=0$).

For any process and any system, equation ( \ref{testqdq2}) can be used as a
test for irreversibility.

It seems a pity, that the concept of the Topological Torsion vector and its
association with non-equilibrium systems, where it can be used to establish
design criteria to minimize energy dissipation, has been ignored by the
engineering community. \ 

\subsubsection{The Spinor class}

It is rather remarkable (and only fully appreciated by me in February, 2005)
that there is a large class of direction fields useful to the topological
dynamics of thermodynamic systems (given herein the symbol\ $\rho
\mathbf{S}_{4})$ that do not behave as vectors (with respect to rotations).
\ They are isotropic complex vectors of zero length, defined as Spinors by E.
Cartan \cite{Cartanspinors}, but which are most easily recognized as the eigen
directionfields relative to the antisymmetric matrix, $[F]$, generated by the
component of the 2-form $F=dA$:%

\begin{align}
\text{\textbf{The Spinor Class \ \ \ }}[F]\circ\left\vert \rho\mathbf{S}%
_{4}\right\rangle  & =\lambda\left\vert \rho\mathbf{S}_{4}\right\rangle
\neq0,\\
\left\langle \rho\mathbf{S}_{4}\right\vert \circ\left\vert \rho\mathbf{S}%
_{4}\right\rangle  & =0,\ \ \ \lambda\neq0
\end{align}
In the language of exterior differential forms, if the Work 1-form is not
zero, the process must contain Spinor components:
\begin{equation}
W=i(\rho\mathbf{S}_{4})dA\neq0
\end{equation}

\ As mentioned above, Spinors have metric properties, behave as vectors with
respect to transitive maps, but do not behave as vectors with respect to
rotations (see p. 3, \cite{Cartanspinors}). \ Spinors generate harmonic forms
and are related to conjugate pairs of minimal surfaces. \ The notation that a
Spinor is a complex isotropic \textit{directionfield} is preferred over the
names "complex isotropic \textit{vector}", or "null \textit{vector}" that
appear in the literature. \ As shown below, the familiar formats of
Hamiltonian mechanical systems exclude the concept of Spinor process
directionfields, for the processes permitted are restricted to be represented
by direction fields of the extremal class, which have zero eigenvalues.

\begin{remark}
Spinors are normal consequences of antisymmetric matrices, and, as topological
artifacts, they are not restricted to physical microscopic or quantum
constraints. \ According to the topological thermodynamic arguments, Spinors
are implicitly involved in all processes for which the 1-form of thermodynamic
Work is not zero.\ \ Spinors play a role in topological fluctuations and
irreversible processes.
\end{remark}

The thermodynamic Work 1-form, $W$, is generated by a completely antisymmetric
2-form, $F$, and therefore, if not zero, must have Spinor components. \ In the
odd dimensional Contact manifold case there is one eigen Vector, with
eigenvalue zero, which generates the extremal processes that can be associated
with a Hamiltonian representation. \ The other two eigendirection fields are
Spinors. \ In the even dimensional Symplectic manifold case, any non-zero
component of work requires that the evolutionary directionfields must contain
Spinor components. \ \ All eigen directionfields on symplectic spaces are Spinors.

\ The fundamental problem of Spinor components is that there can be more than
one Spinor direction field that generates the same geometric path. \ For
example, there can be Spinors of left or right handed polarizations and
Spinors of expansion or contraction that produce the same optical (null
congruence)\ path. \ This result does not fit with the classic arguments of
mechanics, which require unique initial data to yield unique paths.
\ Furthermore, the concept of Spinor processes can annihilate the concept of
time reversal symmetry, inherent in classical hydrodynamics. \ The requirement
of uniqueness is not a requirement of non-equilibrium thermodynamics, where
Spinor "entanglement" has to be taken into account. \ 

\subsection{Emergent Topological Defects}

Suppose an evolutionary process starts in a domain of Pfaff Topological
Dimension 4, for which a process in the direction of the Topological Torsion
vector, $\mathbf{T}_{4}$\ , is known to represent an irreversible process.
\ Examples can demonstrate that the irreversible process can proceed to a
domain of the geometric variety for which the dissipation coefficient,
$\sigma$, becomes zero. \ Physical examples\ \cite{vol2}\ such as the skidding
bowling ball proceed with irreversible dissipation ($PTD=6$) until the
"no-slip" condition is reached ($PTD=5).$ \ In fluid systems the topological
defects\ can emerge as long lived states far from equilibrium. \ The process
is most simply visualized as a "condensation" from a turbulent gas, such as
the creation of a star in the model which presumes the universe is a very
dilute, turbulent van der Waals gas near its critical point. \ The red spot of
Jupiter, a hurricane, the ionized plasma ring in a nuclear explosion, Falaco
Solitons, the wake behind an aircraft are all exhibitions of the emergence
process to long lived topological structures far from equilibrium. \ It is
most remarkable that the emergence of these experimental defect structures
occurs in finite time. \ 

The idea is that a subdomain of the original system of Pfaff Topological
Dimension 4 can evolve continuously with a change of topology to a region of
Pfaff Topological Dimension 3. \ The emergent subdomain of Pfaff Topological
Dimension 3 is a topological defect, with topological coherence, and often
with an extended lifetime (as a soliton structure with a dominant Hamiltonian
evolutionary path), embedded in the Pfaff dimension 4 turbulent background.

The Topological Torsion vector in a region of Pfaff Topological Dimension 3 is
an extremal vector direction field in systems of Pfaff Topological Dimension
3; \ it then has a zero 4D divergence, and leaves the volume element
invariant. \ Moreover the existence of an extremal direction field implies
that the 1-form of Action can be given a Hamiltonian representation,
$P_{k}dq^{k}+H(P,q,t)dt$. \ In the domain of Pfaff dimension 3 for the Action,
$A$, the subsequent continuous evolution of the system, $A$, relative to the
process $\mathbf{T}_{4},$ can proceed in an energy conserving, Hamiltonian
manner, representing a "stationary" or "excited" state far from equilibrium
(the ground state). \ This argument is based on the assumption that the
Hamiltonian component of the direction field is dominant, and any Spinor
components in the $PTD=3$ domain, representing topological fluctuations, can
be ignored.\ \ \ These excited states, far from equilibrium, can be
interpreted as the evolutionary topological defects that emerge and
self-organize due to irreversible processes in the turbulent dissipative
system of Pfaff dimension 4. \ 

The descriptive words of self-organized states far from equilibrium have been
abstracted from the intuition and conjectures of I. Prigogine \cite{Prig}.
\ The methods of Continuous Topological Evolution correct the Prigogine
conjecture that "dissipative structures" can be caused by dissipative
processes and fluctuations. \ The long-lived excited state structures created
by irreversible processes are non-equilibrium, deformable topological defects
almost void of irreversible dissipation. The topological theory presented
herein presents for the first time a solid, formal,\ mathematical
justification (with examples) for the Prigogine conjectures. \ Precise
definitions of equilibrium and non-equilibrium systems, as well as reversible
and irreversible processes can be made in terms of the topological features of
Cartan's exterior calculus. Using Cartan's methods of exterior differential
systems, thermodynamic irreversibility and the arrow of time can be well
defined in a topological sense, a technique that goes beyond (and
without)\ statistical analysis \cite{rmkarw}. \ Thermodynamic irreversibility
and the arrow of time requires that the evolutionary process produce
topological change.

\newpage

\section{Applications}

\subsection{An Electromagnetic format}

The thermodynamic identification of the terms in Cartan's magic formula are
not whimsical. \ To establish an initial level of credence in the terminology,
consider the 1-form of Action, $A$, where the component functions are the
symbols representing the familiar vector and scalar potentials in
electromagnetic theory. \ The coefficient functions have arguments over the
four independent variables $\{x,y,z,t\},$%
\begin{equation}
A=A_{\mu}(x,y,z,t)dx^{\mu}=\mathbf{A}\circ d\mathbf{r}-\phi\ dt.
\end{equation}
Construct the 2-form of field intensities as the exterior differential of the
1-form of Action,%
\begin{align}
F  & =dA=(\partial A_{k}/\partial x^{j}-\partial A_{j}/\partial x^{k}%
)dx^{j\symbol{94}}dx^{k}\\
& =F_{jk}dx^{j}\symbol{94}dx^{k}=+\mathbf{B}_{z}dx\symbol{94}dy...+\mathbf{E}%
_{x}dx\symbol{94}dt...\ \ .
\end{align}
The engineering variables are defined as electric and magnetic field intensities:%

\begin{equation}
\mathbf{E=-\partial A/\partial}t-grad~\phi,\text{ \ \ \ \ \ }\mathbf{B}%
=curl~\mathbf{A.}%
\end{equation}

Relative to the ordered set of base variables, $\{x,y,z,t\}$, define a process
directionfield, $\rho\mathbf{V}_{4}$, as a 4-vector with components,
$[\mathbf{V},1],$ with a scaling factor, $\rho$. \
\begin{equation}
\rho\lbrack\mathbf{V}_{4}]=\rho\lbrack\mathbf{V},1].
\end{equation}
Note that this direction field can be used to construct a useful 3-form of
(matter) current, $C$, in terms of the 4-volume element, $\Omega
_{4}=dx\symbol{94}dy\symbol{94}dz\symbol{94}dt:$
\begin{equation}
C=i(\rho\mathbf{V}_{4})dx\symbol{94}dy\symbol{94}dz\symbol{94}dt=i(\mathbf{C}%
_{4})\Omega_{4}.
\end{equation}
\newline The process 3-form, $C$, is not necessarily the same as
electromagnetic charge current density 3-form of electromagnetic theory, $J$.
\ The 4-divergence of $C$, need not be zero: $\ dC\neq0$. \ 

Using the above expressions, the evaluation of the thermodynamic work 1-form
in terms of 3-vector engineering components becomes:%
\begin{align}
\text{\textbf{The thermodynamic Work 1-form}: }  & W=i(\rho\mathbf{V}%
_{4})dA=i(\rho\mathbf{V}_{4})F,\\
& \Rightarrow-\rho\{\mathbf{E+V\times B}\}\circ d\mathbf{r+\ }\rho
\mathbf{\{V\circ E\}}dt\\
& =-\rho\{\mathbf{f}_{Lorentz}\}\circ d\mathbf{r+\ }\rho\mathbf{\{V\circ
E\}}dt.\\
\text{\textbf{The Lorentz force }}  & =-\{\mathbf{f}_{Lorentz}\}\circ
d\mathbf{r}\text{ (spatial component)}\\
\text{\textbf{The dissipative power}}  & =\mathbf{+\{V\circ E}\}dt\text{ (time
component).}%
\end{align}
\newline

For those with experience in electromagnetism, note that the construction
yields the format, automatically and naturally, for the "Lorentz force" as a
\textit{derivation consequence}, without further ad hoc assumptions. \ The dot
product of a 3 component force, $\mathbf{f}_{Lorentz}$, and a differential
spatial displacement, $d\mathbf{r}$, defines the elementary classic concept of
"differential work". \ \ The 4-component thermodynamic Work 1-form, $W$,
includes the spatial component and a differential time component, $Pdt$, with
a coefficient which is recognized to be the "dissipative power",
$P=\mathbf{\{V\circ E\}}$. \ The thermodynamic Work 1-form, $W$, is not
necessarily a perfect differential, and therefore can be path dependent.
\ Closed cycles of Work need not be zero. \ 

Next compute the Internal Energy term, relative to the process defined as
$\rho\mathbf{V}_{4}$:
\begin{equation}
\text{\textbf{Internal Energy:}}\mathbf{\ }U=i(\rho\mathbf{V}_{4}%
)A=\rho(\mathbf{V\circ A}-\phi).
\end{equation}

The result is to be recognized as the "interaction" energy density in
electromagnetic plasma systems. \ It is apparent that the internal energy, $U
$, corresponds to the interaction energy of the physical system; that is, $U$
is the internal stress energy of system deformation. \ Therefore, the
electromagnetic terminology can be used to demonstrate the premise that
Cartan's magic formula is not just another way to state that the first law of
thermodynamics makes practical sense. \ The topological methods permit the
long sought for integration of mechanical and thermodynamic concepts, without
the constraints of equilibrium systems, and/or statistical analysis.

It is remarkable that although the symbols are different, the same basic
constructions and conclusions apply to many classical physical systems. \ The
correspondence so established between the Cartan magic formula acting on a
1-form of Action, and the first law of thermodynamics is taken both literally
and seriously in this essay. \ \ The methods yield explicit constructions for
testing when a process acting on a physical system is irreversible. \ The
methods permit irreversible adiabatic processes to be distinguished from
reversible adiabatic processes, analytically. \ Adiabatic processes need not
be "slow" or quasi-static.

Given any 1-form, $A,\ W,\ $and/or $Q,$ the concept of Pfaff Topological
Dimension (for each of the three 1-forms, $A$, $W$, $Q$) permits separation of
processes and systems into equivalence classes. \ For example, dynamical
processes can be classified in terms of the topological Pfaff dimension of the
thermodynamic Work 1-form, $W$. \ All extremal Hamiltonian systems have a
thermodynamic Work 1-form, $W$, of topological Pfaff dimension of 1, $(dW=0)$.
\ Hamiltonian systems can describe reversible processes in non-equilibrium
systems for which the topological Pfaff dimension is 3. Such systems are
topological defects whose topology is preserved by the Hamiltonian dynamics,
but all processes which preserve topology are reversible. \ In non-equilibrium
systems, topological fluctuations can be associated with Spinors of the
2-form, $F=dA$. \ Even if the dominant component of the process is
Hamiltonian, Spinor fluctuations can cause the system (ultimately) to decay.

\subsubsection{Topological 3-forms and 4-forms in EM format}

Construct the elements of the Pfaff Sequence for the EM\ notation,%

\begin{equation}
\{A,F=dA,A\symbol{94}F,F\symbol{94}F\},
\end{equation}
and note that the algebraic expressions of Topological Torsion, $A\symbol{94}%
F$, can be evaluated in terms of 4-component engineering variables
$\mathbf{T}_{4}$ as:%

\begin{align}
& \text{\textbf{Topological Torsion vector}}\\
\,\,A\symbol{94}F  & =i(\mathbf{T}_{4})\Omega_{4}=i(\mathbf{T}_{4}%
)dx\symbol{94}dy\symbol{94}dz\symbol{94}dt\\
\mathbf{T}_{4}  & =[\mathbf{T},h]=-[\mathbf{E\times A}+\mathbf{B}%
\phi,\mathbf{A\circ B}].
\end{align}
The exterior 3-form, $A\symbol{94}F$, with physical units of ($\hbar
/$unit\_mole$)^{2}$, is not found (usually) in classical discussions of
electromagnetism\footnote{The unit mole number is charge, e, in EM theory.}. \ 

If $\mathbf{T}_{4}$ is used as to define the direction field of a process,
then
\begin{align}
L_{(\mathbf{T}_{4})}A  & =\sigma A,\ \ \ \ \ i(\mathbf{T}_{4}%
)A=0.\ \ \ \ \ \ \ \\
\text{where \ }2\sigma & =\{div_{4}(\mathbf{T}_{4})\}=2(\mathbf{E}%
\circ\mathbf{B}).\
\end{align}
The important (universal) result is that if the acceleration associated with
the direction field, $\mathbf{E}$, is parallel to the vorticity associated
with the direction field, $\mathbf{B}$\textbf{, }then according to the
equations starting with eq. (\ref{PROPT4}) et. seq. the process is
dissipatively irreversible. \ This result establishes the design criteria for
engineering applications to minimize dissipation from turbulent processes. \ 

The Topological Torsion vector has had almost no utilization in applications
of classical electromagnetic theory. \ 

\subsubsection{Topological Torsion quanta}

The 4-form of Topological Parity, $F\symbol{94}F$, can be evaluated in terms
of 4-component engineering variables as:
\begin{align}
& \text{\textbf{Topological Parity}}\nonumber\\
d(A\symbol{94}F)  & =F\symbol{94}F=\{div_{4}(\mathbf{T}_{4})\}\Omega
_{4}=\{2\mathbf{E\circ B}\}\ \Omega_{4}.
\end{align}
This 4-form is also known as the second Poincare Invariant of Electromagnetic Theory.

The fact that $F\symbol{94}F$ need not be zero implies that the Pfaff
Topological Dimension of the 1-form of Action, $A$, must be 4, and therefore
$A$ represents a non-equilibrium Open thermodynamic system. \ Similarly, if
$F\symbol{94}F=0$, but $A\symbol{94}F\neq0$, then the Pfaff Topological
Dimension of the 1-form of Action, $A$, must be 3, and the physical system is
a non-equilibrium Closed thermodynamic system. \ When $F\symbol{94}F=0,$ the
corresponding three-dimensional integral of the closed 3-form, $A\symbol{94}%
F$, when integrated over a closed 3D-cycle, becomes a deRham period integral,
defined as the Torsion quantum. \ In other words, the closed integral of the
(closed)\ 3-form of Topological Torsion becomes a deformation (Hopf) invariant
with\ integral values proportional to the integers.%
\begin{equation}
\text{\textbf{Torsion quantum}}=\iiint_{3D\_cycle}A\symbol{94}F.\text{\ \ }%
\end{equation}

On the other hand, topological evolution and transitions between
"quantized"\ states of Torsion require that the respective Parity 4-form is
are not zero. \ As,%

\begin{equation}
L_{(\mathbf{T}_{4})}\Omega_{4}=d\{i(\mathbf{T}_{4})\Omega_{4}\}=\ (2\sigma
)\ \Omega_{4}=2(\mathbf{E}\circ\mathbf{B})\ \Omega_{4}\neq0,
\end{equation}
it is apparent that the evolution of the differential volume element,
$\Omega_{4}$, depends upon the existence and colinearity of both the electric
field, $\mathbf{E}$, and the magnetic field, $\mathbf{B}$. \ It is here that
contact is made with the phenomenological concept of "4D bulk" viscosity
$=2\sigma$. It is tempting to identify $\sigma^{2}$ with the concept of
entropy production. \ Note that the Topological Torsion directionfield appears
only in non-equilibrium systems. \ These results are universal and can be used
in hydrodynamic systems discussed in that which follows.

\subsection{A Hydrodynamic format}

\subsubsection{The Topological Continuum vs. the Geometrical Continuum}

In many treatments of fluid mechanics the (geometrical) continuum hypothesis
is invoked from the start. \ The idea is "matter" occupies all points of the
space of interest, and that properties of the fluid can be represented by
piecewise continuous functions of space and time, as long as length and time
scales are not too small. \ The problem is that at very small scales, one has
been led to believe the molecular or atomic structure of particles will become
evident, and the "macroscopic" theory will breakdown. \ However, these
problems of scale are geometric issues, important to many applications, but
not pertinent to a topological perspective, where shape and size are
unimportant. \ \ Suppose that the dynamics can be formulated in terms of
topological concepts which are independent from sizes and shapes. \ Then such
a theory of a Topological Continuum would be valid at all scales. \ Such is
the goal of this monograph. \ 

\begin{remark}
\ However, one instance where "scale" many have topological importance is
associated with the example of a surface with a "teeny" hole. \ If the hole,
no matter what its size, has a twisted ear (Moebius band) then the whole
surface is non-orientable, no matter how "small" the hole. \ Could it be that
the world of the quantum is, in effect, that of non-orientable defects
embedded in an otherwise orientable manifold that originally had no such
defects. \ Note further the strong correspondence with Fermions with
non-oriented (half-integer) multiplet ribbons, and Bosons with oriented
(integer) multiplet ribbons of both right and left twists \cite{vol5}.
\end{remark}

As will be developed below, the fundamental equations of exterior differential
systems can lead to field equations in terms of systems of Partial
Differential Equations (PDE's). \ The format of the fundamental theory will be
in terms of objects (exterior differential forms) which, although composed of
algebraic constructions of tensorial\footnote{Relative to diffeomorphisms.}
things, are in a sense scalars (or pseudo scalars) that are homogeneous with
respect to concepts of scale. \ The theory then developed is applicable to
hydrodynamics at all scales, from the microworld to the cosmological arena.
\ The "breakdown" of the continuum model is not relevant. \ The topological
system may consist of many disconnected parts when the system is not in
thermodynamic equilibrium or isolation, and the parts can have topological
obstructions or defects, some of which can be used to construct period
integrals that are topologically "quantized". \ Hence the "quantization" of
the micro-scaled geometric systems can have it genesis in the non-equilibrium
theory of thermodynamics. \ However, from the topological perspective, the
rational topological quantum values can also occur at all scales.

\subsubsection{Topological Hydrodynamics}

The axioms of Topological Thermodynamics are summarized in Section 1.1. \ For
hydrodynamics (or electrodynamics the axioms are essentially the same. \ Just
exchange the word, hydrodynamics (or electrodynamics) for the word,
thermodynamics, in the formats of Section 1.1

By 1969 it had become evident to me that electromagnetism (without geometric
constraints), when written in terms of differential forms, was a topological
theory, and that the concept of dissipation and irreversible processes
required more than that offered by Hamiltonian mechanics.\ \ At that time I
was interested in possible interactions of the gravitational field and the
polarizations of an electromagnetic signal. \ One of the first ideas
discovered about topological electrodynamics was that there existed an
intrinsic transport theorem \cite{rmkintrinsic} that introduced the concept of
what is now called Topological Spin, $A\symbol{94}G,$ into electromagnetic
theory \cite{vol4}. \ As a transport theorem not recognized by classical
electromagnetism, the first publication was as a letter to Physics of Fluids.
\ That started my interest in a topological formulation of fluids.

It was not until 1974 that the Lie differential acting on exterior
differential forms was established as the key to the problem of intrinsically
describing dissipation and the production of topological defects in physical
systems; but methods of visualization of such topological defects in classical
electrodynamics were not known \cite{rmkhamp}. \ It was hoped that something
in the more visible fluid mechanics arena would lend credence to the concepts
of topological defects. \ The first formulations of the PDE's of fluid
dynamics in terms of differential forms and Cartan's Magic formula followed
quickly \cite{rmkinthydro}. \ 

In 1976 it was argued that topological evolution was at the cause of
turbulence in fluid dynamics, and the notion of what is now called Topological
Torsion, $A\symbol{94}F,$ became recognized as an important concept. \ It was
apparent that streamline flow imposed the constraint that $A\symbol{94}F=0$ on
the equations of hydrodynamics. \ Turbulent flow,\ being the antithesis of
streamline flow, must admit $A\symbol{94}F\neq0.$ \ In 1977 it was recognized
that topological defect structures could become "quantized" in terms of deRham
period integrals \cite{rmkperiods}, forming a possible link between topology
and both macroscopic and microscopic quantum physics. \ The research effort
then turned back to a study of topological electrodynamics in terms of the
dual polarized ring laser, where it was experimentally determined that the
speed of an electromagnetic signal outbound could be different from the speed
of an electromagnetic signal inbound: \ a topological result not within the
realm of classical theory.

Then in 1986 the long sought for creation and visualization of topological
defects in fluids \cite{rmkfalaco} became evident. \ The creation of Falaco
Solitons in a swimming pool was the experiment that established credence in
the ideas of what had, by that time, become a theory of continuous topological
evolution. \ It was at the Cambridge conference in 1989 \cite{rmkmoffat} that
the notions of topological evolution, hydrodynamics and thermodynamics were
put together in a rudimentary form, but\ it was a year later at the Permb
conference in 1990 \cite{rmkpermb} that the ideas were well established. \ The
Permb presentation also suggested that the ambiguous (at that time) notion of
coherent structures in fluids could be made precise in terms of topological
coherence. \ A number of conference presentations followed in which the ideas
of continuous thermodynamic irreversible topological evolution in
hydrodynamics were described \cite{RMKsectam}, but the idea that the
topological methods of thermodynamics could be used to distinguish
non-equilibrium processes and non-equilibrium systems and irreversible
processes with out the use of statistics slowly came into being in the period
1985-2005 \cite{rmkvig2000}. \ These efforts have been summarized
in\ \cite{vol1}, and a collection of the old publications appears in
\cite{vol7}.

\subsection{Classical Hydrodynamic Theory}

There are two classical techniques for describing the evolutionary motion of a
fluid: the Lagrangian method and the Eulerian method. \ Both methods treat the
fluid relative to a Euclidean 3D manifold, with time as a parameter. \ The
first (Lagrangian) technique treats a fluid as a collection of "particles, or
parcels" and the flow is computed in terms of "initial" data $\{a,b,c,\tau\}$
imposed upon solutions to Newtonian equations of motion for "particles, or
parcels". \ \ The solution functions describe a map from an initial state
$\{a,b,c;\tau\}~$to a final state $\{x,y,z,t\}$. \ This method is related to
solutions of kinematic equations, and is \textit{contravariant} in the sense
of an immersion to velocities (the tangent space). \ The kinematic basis for
the Lagrangian motion draws heavily from the Frenet-Serret analysis of a point
moving along a space curve.

The second technique treats a fluid as "field", and is representative of a
"wave" point of view of a Hamiltonian system. \ The functions that define the
field (the covariant momenta) depend on "final data" $\{x,y,z,t\},$ and are
covariant in the sense of a submersion. \ Each method has its preimage in the
form of an exterior differential 1-form of Action. \ The primitive classical
Lagrangian Action concept is written in the form
\begin{equation}
A_{L}=\mathsf{L}(x^{k},V^{m},t)dt,
\end{equation}
and the primitive Eulerian Action is written as%

\begin{equation}
A_{E}=p_{k}dx^{k}+\mathsf{H}dt.
\end{equation}
As both Actions supposedly describe the same fluid, are they equivalent?
\ That is,%

\begin{equation}
\text{does }A_{E}\Leftrightarrow A_{L}\text{ ?}%
\end{equation}
Note that $A_{L}$ is composed from only two functions ($\mathsf{L}$ and $t$)
such that at most $A_{L}$ is\ of Pfaff Topological Dimension 2. \ On the other
hand, the Pfaff Topological Dimension of $A_{E}$ (as written) could be as high
as 8, if all functions and differentials are presumed to be independent. \ So
the two formulations are NOT equivalent, unless constraints reduce the
topological dimension of $A_{E}$ to 2, or if additions are made to the 1-form
$A_{L}$ to increase its Pfaff dimension.

\subsubsection{The Lagrange-Hilbert Action}

The classic addition to $A_{L}$ is of the form, \ $p_{k}(dx^{k}-V^{k}dt)$,
where the $p_{k}$ are presumed to be Lagrange multipliers of the fluctuations
in kinematic velocity. \ The result is defined as the Cartan - Hilbert 1-form
of Action:%

\begin{equation}
A_{CH}=\mathsf{L}(x^{k},V^{k},t)dt+p_{k}(dx^{k}-V^{k}dt).
\end{equation}
Note that at first glance it appears that there are 10=3N+1 independent
geometric variables $\{x^{k},V^{k},p_{k},t\}$ in the formula for $A_{CH}$, but
if the Pfaff Sequence is constructed, the Pfaff Topological Dimension turns
out to be 8. \ So with this addition of Lagrange multipliers to $A_{L}, $ the
topological dimensions of the two actions are the same. \ However, note that
by rearranging variables,%

\begin{align}
A_{CH}  & =p_{k}dx^{k}+(\mathsf{L}(x^{k},V^{m},t)-p_{k}V^{k})dt,\\
& =p_{k}dx^{k}+\mathsf{H}dt=A_{E}.
\end{align}
For a fluid the Eulerian "momenta" per unit parcel of mass is usually defined
as
\begin{equation}
p_{k}/m=\mathbf{v}_{k},
\end{equation}
such that the Eulerian Action per unit mass becomes%

\begin{equation}
A_{E}\Rightarrow\mathbf{v}_{k}dx^{k}+\mathsf{H}dt.
\end{equation}
The bottom line is that the Lagrangian and Eulerian point of view can be made
compatible if fluctuations in Kinematic Perfection are allowed. \ 

Recall that the development of elasticity theory (and its emphasis on
symmetrical tensors) also spawned the development of hydrodynamics. \ Much of
the theory of classical hydrodynamics was phrased geometrically in the
language of vector analysis. \ The development followed the phenomenological
concepts of an extended Newtonian theory of elasticity. \ The classical theory
was developed from "balance" equations for a bounded sample, or parcel, of
matter (mass), which express assumptions (defined as the conservation of mass,
momenta and energy) in terms of integrals over the bounded sample, or parcel,
of matter. \ The classical integrals are performed usually over
three-dimensional volumes (and not over 4D space-time). \ The classical Cauchy
result (for the momentum equations) is%

\begin{align}
\rho\{\partial\mathbf{v}/\partial t+\mathbf{v\circ\nabla v\}}  &
=\mathbf{\nabla\circ}\mathbb{T}+\rho\mathbf{f,}\label{cauchy}\\
\rho\{\partial\mathbf{v}/\partial t+grad(\mathbf{v\circ v/2)-v\times
}curl\mathbf{v\}}  & =\mathbf{\nabla\circ}\mathbb{T}+\rho\mathbf{f,}%
\end{align}
Constitutive assumptions are then made for the 3D\ stress tensor
$\mathbb{T}\mathbf{\,}$, such that (in matrix format)$\ $%
\begin{equation}
\left[  \mathbb{T}\right]  =(-P+\lambda(\mathbf{\nabla\circ v})\left[
\mathbb{I}\right]  +\nu\{[\mathbf{\nabla v}]+[\mathbf{\nabla v}]^{T}\},
\end{equation}
where $P$ is the Pressure, $\nu$ the affine "shear" viscosity and $\lambda$
the "expansion" viscosity. \ The antisymmetric components $\{[\mathbf{\nabla
v}]-[\mathbf{\nabla v}]^{T}\}$ have been ignored. \ 

It will be demonstrated how the Axioms of Hydrodynamics yield topological
information about the classic Cauchy development, and goes beyond the
symmetrical formulations by recognizing that antisymmetries can introduce
complex spinor contributions to the dynamics.

\subsection{Euler flows and Hamiltonian fluids}

Consider the Action 1-form per unit source (in thermodynamics, the unit source
is mole number, or sometimes mass), constructed from a covariant 3D velocity
field, $\mathbf{v=v}_{k}(x,y,z,t)$, and a scalar potential function, $\phi$:
\begin{equation}
A=\mathbf{v\circ dr-}\phi dt=\mathbf{v}_{k}(x,y,z,t)dx^{k}-\phi dt.
\end{equation}
\ Compute the exterior differential $dA$ and define the following (3D vector)
functions as,
\begin{equation}
\mathbf{\omega}=curl\ \mathbf{v\ \ \ \ }and\ \ \ \mathbf{a}=+\{\partial
\mathbf{v}/\partial t+grad(\phi)\},
\end{equation}
such that,%
\begin{align}
F &  =dA=\{\partial A_{k}/\partial x^{j}-\partial A_{j}/\partial x^{k}%
\}dx^{j}\symbol{94}dx^{k}=F_{jk}dx^{j}\symbol{94}dx^{k}\\
&  =\mathbf{\omega}_{z}dx\symbol{94}dy+\mathbf{\omega}_{x}dy\symbol{94}%
dz+\mathbf{\omega}_{y}dz\symbol{94}dx-\mathbf{a}_{x}dx\symbol{94}%
dt-\mathbf{a}_{y}dy\symbol{94}dt-\mathbf{a}_{z}dz\symbol{94}dt.\nonumber
\end{align}

These vector fields always satisfy the Poincare-Faraday induction equations,
$dF=ddA=0,$ or$,$%
\begin{equation}
curl\ (-\mathbf{a)}+\partial\mathbf{\omega}/\partial
t=0,\ \ \ \ div\ \mathbf{\omega}=0.
\end{equation}

\paragraph{The Eulerian Fluid}

Consider a process created by the contravariant vector directionfield,
$\mathbf{V}_{4}=$ $[\mathbf{V}^{x},\mathbf{V}^{y},\mathbf{V}^{z},1]$ and use
Cartan's magic formula,%

\begin{equation}
L_{(\rho\mathbf{V}_{4})}A=i(\rho\mathbf{V}_{4})dA+d(i(\rho\mathbf{V}%
_{4})A)=W+dU=Q,
\end{equation}
to compute the thermodynamic Work 1-form, $W$. \ The expressions for Work,
$W,$ and internal energy, U, become:%
\begin{align}
W  & =i(\rho\mathbf{V}_{4})dA=-\rho\{-\partial\mathbf{v}/\partial
t-grad(\phi)+\mathbf{V\times\omega\}}\circ d\mathbf{r}\nonumber\\
& -\rho\mathbf{V}\circ\{\partial\mathbf{v}/\partial t+grad(\phi)\}dt,\\
& =\rho\{\mathbf{a}-\mathbf{V\times\omega\}}\circ d\mathbf{r-}\rho
\mathbf{V}\circ\{\mathbf{a})\}dt\\
U  & =i(\mathbf{V}_{4})A=\rho(\mathbf{V\cdot v-}\phi).
\end{align}

At first, topologically constrain the thermodynamic Work 1-form to be of the
Bernoulli class in terms of the exterior differential system:%
\begin{align}
W  & =-dP\\
\rho\{+\mathbf{a}-\mathbf{V\times\omega\}}\circ d\mathbf{r-}\rho
\mathbf{V}\circ\{\mathbf{a})\}dt  & =-grad~P\circ d\mathbf{r-}\partial
P/\partial t~dt
\end{align}
\ Assume that formally $\mathbf{V=v}$, and the potential is equal to
$\phi=\mathbf{v\cdot v}/2$. \ Compare the coefficients of $d\mathbf{r}$ to
deduce the classic equations of motion for the Eulerian fluid.%
\begin{equation}
\{\partial\mathbf{v}/\partial t+grad(\mathbf{v\cdot v}/2)-\mathbf{v\times
\omega\}}=-grad(P)/\rho.\label{LEfluid}%
\end{equation}

This formula should be compared to the derivation of the Lorentz force term
for Work in electromagnetic systems. \ The functional format of the
hydrodynamic 1-form of Action, $A$, is the same as that specified above for
the electromagnetic system. \ All that is changed is the notation. \ In
essence, the two topological theories are equivalent to the extent that there
is a correspondence between functions:%

\begin{align}
\mathbf{A}  & \Leftrightarrow\mathbf{v,\ \ \ \ \ \ \ }\phi\Leftrightarrow
\mathbf{v\cdot v}/2,\\
\mathbf{E}  & \Leftrightarrow-\mathbf{a,\ \ \ \ B\Leftrightarrow\omega
}\text{.}%
\end{align}
All the results of the preceding section using an electromagnetic format can
be translated into the hydrodynamic format. \ 

Note that the Bernoulli "pressure", $P$, is an evolutionary invariant along a trajectory,%

\begin{equation}
L_{(\rho\mathbf{V}_{4})}P=i(\mathbf{V}_{4})dP=i(\mathbf{V}_{4})i(\mathbf{V}%
_{4})dA=0.
\end{equation}
If the Pressure is barotropic, then the Bernoulli function becomes,
$d\Theta=dP/\rho.$ \ The function $\Theta$ can be amalgamated with the
potential, $\phi$, such that the thermodynamic Work 1-form becomes equal to
zero. \ The system then admits an extremal Hamiltonian direction field such
that the thermodynamic Work 1-form is zero.%

\begin{equation}
W=i(V_{H})dA=0.
\end{equation}
For a process defined in terms of an extremal directionfield, the First Law
indicates that the 1-form of Heat, $Q,$ is exact, $dQ=0$, and equal to the
change in internal energy, $Q=dU$. \ Any Hamiltonian process is reversible, as
$Q\symbol{94}dQ=0$.

The time-like component of the exterior differential system $W+dP=0$ leads to
the equation,%

\begin{equation}
\partial P/\partial t=-\rho\mathbf{v}\circ\{\partial\mathbf{v}/\partial
t+grad(\mathbf{v\cdot v}/2)=\rho(\mathbf{v}\circ\mathbf{a)}.
\end{equation}
It is apparent that if the velocity, $\mathbf{v}$, and the acceleration,
$\mathbf{a}$, are orthogonal, then the time rate of change of the Bernoulli
pressure is zero.

It also follows that the "Master" equation is valid, with the only difference
being that $curl\ \mathbf{v}$ is defined as $\mathbf{\omega}$, the vorticity
of the hydrodynamic flow. \ The master equation becomes,%
\begin{equation}
curl(\mathbf{v\times\omega})=\partial\mathbf{\omega}/\partial t,
\end{equation}
and this equation is to be recognized as the equivalent of Helmholtz' equation
for the conservation of vorticity. \ When the Pfaff Topological Dimension of
the Work 1-form is 1, it is possible to show that the "Master" leads to a
diffusion equation. \ 

In the hydrodynamic sense, conservation of vorticity implies uniform
continuity. \ In other words, the Eulerian flow is not only Hamiltonian, it is
also uniformly continuous, and satisfies both the master equation and the
conservation of vorticity constraints. In addition, it may be demonstrated
that such systems are at most of Pfaff dimension 3, and admit a relative
integral invariant which generalizes the hydrodynamic concept of invariant
helicity. In the electromagnetic topology, the Hamiltonian constraint is
equivalent to the statement that the Lorentz force vanishes, a condition that
has been used to define the "ideal"\ plasma or "force-free"\ plasma state
\cite{Wesson}.

\newpage

\section{The Navier-Stokes fluid}

\subsection{The classic Navier-Stokes equations}

It can be demonstrated that the "ideal fluid" has a Hamiltonian
representation, for which the dynamics preserves a "Hamiltonian" energy.
\ This result is in disagreement with experiment in that it is observed that
motions of "non-ideal" fluids exhibit decay to a stationary state. \ The
Lagrange Euler equations must be modified to accommodate dissipation of
kinetic energy and angular momentum. \ In fact, the ideal fluid constraint of
zero affine shear stresses should be replaced by dissipative terms related to
both affine shears and a new phenomena of rotational and expansion shears
which have a fixed point. \ The classical phenomenological outcome is the
Navier-Stokes PDE's,%

\begin{align}
\partial\mathbf{v}/\partial t+grad(\mathbf{v\cdot v}/2)-\mathbf{v\times
}\ curl\ \mathbf{v}  & =-grad\varphi-(1/\rho)\ gradP\nonumber\\
& +\nu\Delta\mathbf{v}\nonumber\\
& -(\mu_{B}+\nu)\ grad\ div\ \mathbf{v},\label{NSFORMAT}%
\end{align}
where $\mu_{B}~$is the "bulk" viscosity coefficient and $\nu$ is the "shear"
viscosity coefficient. \ If the fluid is "incompressible" then the last term,
which includes corrections due to bulk viscosity, vanishes; \ the
incompressible constraint requires that $div\ \mathbf{v}\Rightarrow
0\mathbf{.}$

In that which follows, the basic momentum equation (\ref{NSFORMAT}) will be
\textit{deduced} from the perspective of Continuous Topological Evolution.
\ Different topological equivalence classes of thermodynamic processes depend
upon the Pfaff Topological Dimension of the Work 1-form. \ The different
classes of thermodynamic processes are related to the velocity field in a
Hydrodynamic system. \ The phenomenological (geometrical) derivation of the
equations of hydrodynamics will be replaced by determining the format of the
PDE's that agree with the constraint required to satisfy the various PTD
equivalence classes the Work 1-form. \ The 1-form of Work (for barotropic
flows as in eq. (\ref{LEfluid})) will be of Pfaff Topological Dimension 1.
\ The Pfaff Topological Dimension of the 1-form of Work for the Navier-Stokes
fluid can be as high as 4, and is required to be 4 if the flow is fully
turbulent. \ Various intermediate classes of the work 1-form are of interest,
as well. \ In particular, the Pfaff Topological Dimension of the work 1-form
must be 3 for a baroclinic system, a result that admits frontal systems with
propagating tangential discontinuities as found in weather systems.

\subsection{The Navier-Stokes equations embedded in a non-equilibrium
thermodynamic system}

In this subsection, the topological refinement due to the Pfaff Topological
Dimension of the Work 1-form will be employed to demonstrate that processes in
a non-equilibrium thermodynamic system can be put into correspondence with
solutions of the Navier-Stokes equations.

In order to go beyond extremal (Hamiltonian) processes, it is necessary that
the Pfaff Topological Dimension of the Work 1-form must be greater than 1.
\ Recall that for any process, the Work done is transverse to the process trajectory,%

\begin{equation}
(i(\rho\mathbf{V}_{4})W=(i(\rho\mathbf{V}_{4})(i(\rho\mathbf{V}_{4})dA=0.
\end{equation}
Hence, if the PTD\ of the Work 1-form, $W$, is to be greater than 1, it must
have the format,
\begin{equation}
W=i(\rho\mathbf{V}_{4})dA=-dP+\varpi_{j}(d\mathbf{x}^{j}-\mathbf{v}%
^{j}dt)=-dP+\varpi_{j}\Delta\mathbf{x}^{j},
\end{equation}
where the "Bernoulli function", $P$, if it exists, must be a first integral (a
process invariant),%

\begin{equation}
L(\rho\mathbf{V}_{4})P=(i(\rho\mathbf{V}_{4})dP=0.
\end{equation}

It is also important to remember that such non-zero contributions to the work
1-form are due to the complex, isotropic Cartan Spinors, which are the eigen
direction fields of the 2-form, $F$.

The coefficients, $\varpi_{j}$, of the topological fluctuations,
$\Delta\mathbf{x}^{j}$, act in the manner of Lagrange multipliers, and mimic
the concept of system forces. \ If $\varpi_{j}/\rho$ is defined
(arbitrarily\footnote{This is one of many formal choices, but the choice
demonstrates that the Navier-Stokes equations reside within the domain of
non-equilibrium thermodynamics. QED}) as $\upsilon\,curl\,curl\,\mathbf{v}$
then the spatial components of the thermodynamic Work 1-form, $W$, are
constrained to yield the partial differential equations for a constant density
Navier-Stokes fluid:%
\begin{equation}
\{\partial\mathbf{v}/\partial t+grad(\mathbf{v\cdot v}/2)-\mathbf{v\times
\omega\}}=-grad(P)/\rho+\upsilon\,curl\,curl\,\mathbf{v.}%
\end{equation}
Density variations can be included by adding a term $\lambda div(\mathbf{V})$
to the potential $\{\mathbf{v\cdot v}/2\}$ to yield:%
\begin{align}
\partial\mathbf{v}/\partial t+grad\{\mathbf{v.v}/2\}-\mathbf{v}\times
curl\ \mathbf{v}  & =-gradP/\rho\\
& +\lambda\{grad(div~\mathbf{v)\}}\label{NavStokes}\\
& +\upsilon\{curl\ curl\ \mathbf{v}\}.
\end{align}
Classically, $v$ can be identified with the geometric kinematic shear
viscosity, and $\lambda=\mu_{B}-\upsilon.$ \ The coefficient $\mu_{B}$ can be
identified with the topological (space-time) bulk viscosity.\ 

\subsection{The Topological Torsion process for the Navier-Stokes fluid}

The Navier-Stokes constraint implies that the thermodynamic Work 1-form need
not be closed. \ Then there are thermodynamic processes represented by
solutions to the Navier-Stokes equations that are thermodynamically
irreversible. \ In this subsection, the Topological Torsion vector will be
expressed in terms of the solutions of the Navier-Stokes equations.\ 

From the work in section 2, the 1-form of Action will generate a 3-form of
Topological Torsion, $A\symbol{94}dA=i(\mathbf{T}_{4})\Omega_{4},$ and leads
to the 4 Vector of Topological Torsion (written in hydrodynamic notation):%
\begin{align}
\mathbf{T}_{4}  & =[-\mathbf{a\times v}+\{\mathbf{v.v}%
/2\}\ curl\ \mathbf{v,(v}\circ curl\,\mathbf{v})],\\
& =[-\mathbf{a\times v}+\{\mathbf{v.v}/2\}\ \mathbf{\omega,}(\mathbf{v}%
\circ\mathbf{\omega})]=[\mathbf{T},h].
\end{align}
Use the Navier-Stokes equations (\ref{NSFORMAT}) to solve for $\mathbf{a,}$%
\begin{align}
\mathbf{a}  & =[grad\{\mathbf{v.v}/2\}+\partial\mathbf{v}/\partial t]\\
& =\mathbf{v}\times curl\ \mathbf{v}-gradP/\rho\nonumber\\
& +\lambda\{grad(div~\mathbf{v)\}}+\upsilon\{curl\ curl\ \mathbf{v}\},
\end{align}
and then substitute this result into the expression for $\mathbf{T}_{4},$ to
yield:%
\begin{align}
\mathbf{T}  & =[h\mathbf{v}-(\mathbf{v\circ v}/2)curl~\mathbf{v}%
-\mathbf{v}\times(gradPl\rho)\nonumber\\
& +\lambda\{\mathbf{v}\times grad(div~\mathbf{v)\}}-\upsilon\{\mathbf{v}%
\times(curl\ curl\ \mathbf{v})\}],\\
h  & =\mathbf{v}\cdot curl\,\mathbf{v,}%
\end{align}
Note that $\mathbf{T}_{4}$ exists even for Euler flows, where $\upsilon=0,$ if
the flow is baroclinic. \ The measurement of the components of the Torsion
vector, $\mathbf{T}_{4},$ have been completely ignored by experimentalists in hydrodynamics.

By a similar substitution, the topological parity 4-form, $F\symbol{94}F,$
becomes expressible in terms of engineering quantities as,%
\begin{align}
K  & =\{2(-\mathbf{a}\circ\mathbf{\omega})\}\Omega_{4}=\{2(\sigma
)\}\Leftrightarrow2(\mathbf{E\circ B)}\nonumber\\
\sigma & =\{gradP/\rho\circ curl\ \mathbf{v}\\
& -\lambda\{\ grad(div~\mathbf{v)}\circ curl\ \mathbf{v\}}\nonumber\\
& -\upsilon\{curl\mathbf{\ v}\circ(curl\ curl\ \mathbf{v})\}\}\Omega_{4}.
\end{align}
The coefficient $\sigma$ is a measure of the space-time bulk dissipation
coefficient (not $\lambda$), and it is the square of this number which must
not be zero if the process is irreversible (see eq (\ref{testqdq2})). \ Recall
that a turbulent dissipative irreversible flow is defined when the Pfaff
dimension of the Action 1-form is equal to 4, which implies that $K\neq0.$ \ 

From the expression for $\sigma$, it is apparent that if the 3D vector of
vorticity is of Pfaff dimension 2, such that $\mathbf{\omega\circ
}curl\ \mathbf{\omega}=0$, then the last term vanishes, and there is no
irreversible dissipation due to shear viscosity, $\upsilon$ (a result useful
in the theory of wakes). \ 

Other useful situations and design criteria for dissipation, or the lack
thereof, can be gleaned from the formula. \ If the vector field is harmonic,
then an irreversible process requires that,%
\begin{equation}
\sigma=(-\mathbf{a}\circ\mathbf{\omega})=\{(gradP/\rho-\mu_{B}%
grad(div~\mathbf{v))}\circ curl\ \mathbf{v\}}\neq0.
\end{equation}
(Recall that harmonic vector fields are generators of minimal surfaces.) \ For
fluids where $(\mu_{B})\Rightarrow0,$ if the pressure gradient is orthogonal
to the vorticity and the flow field is harmonic, then there is no irreversible
dissipation as $\sigma=0$, and the flow is not turbulent. \ Note that for many
fluids the bulk viscosity is much greater than the shear viscosity. \ When
$\sigma=0$, no topological torsion defects are created; \ the acceleration,
$\mathbf{a}$, and the vorticity, $\mathbf{\omega}$, of the Navier-Stokes fluid
are colinear. \ 

\begin{theorem}
It is thereby demonstrated that solutions to the Navier-Stokes equations
correspond to processes of a non-equilibrium thermodynamic system of
$PTD(A)=4$, and Work 1-forms of $PTD(W)>2.$ \ Such processes include Spinor
direction fields generated by the Topological Torsion vector. \ The
Topological Torsion vector generates processes, and hence solutions to the
Navier-Stokes equations, that are thermodynamically irreversible..\ 
\end{theorem}

These results should be compared to those generated by Lamb and Eckart
\cite{Eckart} for the fluid dissipation function, which is defined by the
requirement that the dissipative flow has a (geometric) entropy production
rate greater than or equal to zero. \ More examples can be found in, "Wakes,
Coherent Structures, and Turbulence" \cite{vol3}.

\newpage

\section{Closed States of Topological Coherence embedded as deformable defects
in Turbulent Domains}

In this section 4, the problem of C2 smoothness will be attacked from the
point of view of topological thermodynamics. \ First, two distinct examples
will be given demonstrating two different emergent $PTD=3$ states, that emerge
from different 4D\ rotations (see p. 108, \cite{Stewart}). \ Then, an example
demonstrating the decay of a $PTD=4$ state into a $PTD=3$ state will be given
in detail.

\subsection{Examples of PTD = 3 domains and their Emergence}

\ In section 3, it was demonstrated that there are solutions (thermodynamic
processes) to the Navier Stokes equations in non-equilibrium thermodynamic
domains. \ The properties of those $PTD=3$ domains which emerge by C2
irreversible solutions from domains of $PTD=4$ are of particular interest.
\ From section 2, it is apparent that the key feature of $PTD=3$ domains is
that the electric $\mathbf{E}$ field (acceleration\ field $\mathbf{a}%
$\textbf{\ }in hydrodynamics) must be orthogonal to the magnetic $\mathbf{B}$
field (vorticity field $\mathbf{\omega}$ in hydrodynamics). \ There are 8
cases to consider (including chirality), \
\begin{align}
& \text{\textbf{Pfaff Topological Dimension 3}}\nonumber\\
\mathbf{E\ }\mathbf{=0,}  & \text{ \ \ \ }\mathbf{\pm B\neq0,\ \ }\\
\text{\ \ }\mathbf{B\ }\mathbf{=0,}  & \text{ \ \ \ }\mathbf{\pm E\neq0,}\\
\mathbf{E\circ B\ }\mathbf{=0,}  & \text{ with chirality choices, }%
\pm\mathbf{E=\pm B\neq0,}%
\end{align}
of which two will be discussed in detail.

\paragraph{The Finite Helicity case (both E and B finite) PTD = 3}

Start with the 4D thermodynamic domain, and first consider the 1-form of
Action, $A$, with the format\footnote{The +E, +B chirality has been selected.}:%

\begin{equation}
A=A_{x}(z)dx+A_{y}(z)dy-\phi(z)dt,
\end{equation}
and its induced 2-form, $F=dA$,%
\begin{align}
F  & =dA=(\partial A_{x}(z)/\partial z)dz\symbol{94}dx+(\partial
A_{y}(z)/\partial z)dz\symbol{94}dy-(\partial\phi(z)/\partial z)dz\symbol{94}%
dt,\\
& =B_{x}(z)dz\symbol{94}dx-B_{y}(z)dz\symbol{94}dy+E_{z}(z)dz\symbol{94}dt.
\end{align}
The 3-form of Topological Torsion 3-form becomes%
\begin{align}
i(\mathbf{T}_{4})\Omega_{4}  & =A\symbol{94}F\text{ where}\\
\mathbf{T}_{4}(z)  & =[E_{z}A_{y}+\phi B_{x},~-E_{z}A_{x}+\phi B_{y}%
,\ 0,\ A_{x}B_{x}+A_{y}B_{y}]\\
\text{with }div_{4}(\mathbf{T}_{4}(z))  & =2(\mathbf{E\circ B)=\ }%
0,\ \ \ \ \ \ \mathbf{A\circ B}\neq0.
\end{align}

\paragraph{The Zero Helicity case (both E and B finite) PTD = 3}

Start with the 4D thermodynamic domain, and consider the 1-form of Action, $A
$, with the format:%

\begin{equation}
A=A_{x}(x,y)dx+A_{y}(x,y)dy-\phi(x,y)dt,
\end{equation}
and its induced 2-form, $F=dA$,%
\begin{align}
F  & =dA=\{(\partial A_{y}(x,y)/\partial x)-(\partial A_{x}(x,y)/\partial
x)dx\symbol{94}dy\}\nonumber\\
& -(\partial\phi(x,y)/\partial x)dx\symbol{94}dt-(\partial\phi(x,y)/\partial
y)dy\symbol{94}dt,\\
& =B_{z}(x,y)dx\symbol{94}dy+E_{x}(x,y)dx\symbol{94}dt+E_{y}(x,y)dy\symbol{94}%
dt.
\end{align}
The 3-form of Topological Torsion 3-form becomes,%
\begin{align}
i(\mathbf{T}_{4})\Omega_{4}  & =A\symbol{94}F\text{ \ \ \ \ where}\\
\mathbf{T}_{4}(x,y)  & =[0,0,(E_{x}A_{y}-E_{y}A_{x})+\phi B_{z},0]\\
\text{with }div_{4}(\mathbf{T}_{4}(x,y))  & =2(\mathbf{E\circ B)=\ }%
0,\ \ \ \ \mathbf{A\circ B}=0.
\end{align}
This case of zero helicity ($\mathbf{A\circ B=0}$), has the Topological
Torsion vector, $\mathbf{T}_{4}(x,y)$, colinear with the\textbf{\ }%
$\mathbf{B}$ field.

\paragraph{Zero Helicity case: PTD = 4 decays to PTD = 3}

The two distinct cases, modulo chirality, are suggestive of the idea (see p.
108 \cite{Stewart}) that the rotation group of a 4D domain is not simple.
\ The example,immediately above, is particularly useful because the algebra of
the decay from Pfaff dimension 4 to 3 is transparent. \ 

Start with the 4D thermodynamic domain, and consider the 1-form of Action, $A
$, with the format:%

\begin{equation}
A=A_{x}(x,y)dx+A_{y}(x,y)dy-\phi(x,y,z,t)dt,
\end{equation}
and its induced 2-form, $F=dA$,%
\begin{align}
F  & =dA=\{(\partial A_{y}(x,y)/\partial x)-(\partial A_{x}(x,y)/\partial
x)dx\symbol{94}dy\}\nonumber\\
& -(\partial\phi(x,y,z)/\partial x)dx\symbol{94}dt-(\partial\phi
(x,y,z)/\partial y)dy\symbol{94}dt-(\partial\phi(x,y,z)/\partial
z)dz\symbol{94}dt,\\
& =B_{z}(x,y)dx\symbol{94}dy\\
& +E_{x}(x,y,z,t)dx\symbol{94}dt+E_{y}(x,y,z,t)dy\symbol{94}dt+E_{z}%
(x,y,z,t)dy\symbol{94}dt.
\end{align}
The 3-form of Topological Torsion 3-form becomes,%
\begin{align}
i(\mathbf{T}_{4})\Omega_{4}  & =A\symbol{94}F\text{ with PTD(}A\text{) = 4}\\
\text{ \ }\mathbf{T}_{4}(x,y,z,t)  & =[-E_{z}A_{y},+E_{z}A_{x},(E_{x}%
A_{y}-E_{y}A_{x})+\phi B_{z},0]\\
\text{with }div_{4}(\mathbf{T}_{4}(x,y,z,t))  & =2\{E_{z}(x,y,z,t)B_{z}%
(x,y)\}\ \mathbf{\neq\ }0.
\end{align}
In this case, the helicity ($\mathbf{A\circ B=0}$) is still zero, but now the
Topological Torsion vector, $\mathbf{T}_{4}(x,y,z,t)$, has three spatial
components. \ Moreover, the Process generated by\ $\mathbf{T}_{4}(x,y,z)$ is
thermodynamically irreversible, as $(\mathbf{E\circ B)\neq\ }0$. \ The example
1-form is of PTD\ = 4.

To demonstrate the emergence of the $PTD=3$ state, suppose the potential
function in this example has the format,%

\begin{align}
\phi & =\psi(x,y)+\varphi(z)e^{-\alpha t}\\
E_{z}(z,t)  & =-(\partial\varphi(z)/\partial z)e^{-\alpha t}=E_{z}%
(z)e^{-\alpha t}.
\end{align}
Then the irreversible dissipation function decays as $\{E_{z}(z)B_{z}%
\}e^{-\alpha t}$. \ By addition of Spinor fluctuation terms to represent the
very small components of irreversible dissipation at late times, the $PTD=3$ solution,%

\begin{equation}
\mathbf{T}_{4}(x,y)=[0,0,(E_{x}A_{y}-E_{y}A_{x})+\phi B_{z},0]
\end{equation}
becomes dominant, and represents a long lived "stationary" state far from
equilibrium, modulo the small Spinor decay terms\footnote{The experimental
fact that the defect structures emerge in finite time is still an open
\textit{topological} problem, although some \textit{geomtric} success has been
achieved through Ricci flows.}. \ 

\subsection{Piecewise Linear Vector Processes vs. C2 Spinor processes}

It will be demonstrated on thermodynamic spaces of Pfaff Topological Dimension
3, that there exist piecewise continuous processes (solutions to the
Navier-Stokes equations) which are thermodynamically reversible. \ These
Vector processes can be fabricated by combinations of Spinor processes, each
of which is irreversible. \ This topological result demonstrates, by example,
the difference between piecewise linear 3-manifolds and smooth complex
manifolds. \ It appears that the key feature of the irreversible processes is
that they have a fixed point of "rotation or expansion".

Consider those abstract physical systems that are represented by 1-forms, $A$,
of Pfaff Topological Dimension 3. \ The concept implies that the topological
features can be described in terms of 3 functions (of perhaps many geometrical
coordinates and parameters) and their differentials. \ For example, if one
presumes the fundamental independent base variables are the set $\{x,y,z\},$
with an exterior differential oriented volume element consisting of a
product\footnote{More abstract systems could be constructed from differential
forms which are not exact.} of exact 1-forms $\Omega_{3}=+dx\symbol{94}%
dy\symbol{94}dz,$ (then a local) Darboux representation for a physical system
could have the appearance,%
\begin{equation}
A=xdy+dz.
\end{equation}
The objective is to use the features of Cartan's magic formula to compute the
possible evolutionary features of such a system. \ \ The evolutionary dynamics
is essentially the first law of thermodynamics:%
\begin{equation}
L_{\rho\mathbf{V}}A=i(\rho\mathbf{V})dA+di(\rho\mathbf{V)}A)\ \mathbf{=\ }%
W+dU=Q.
\end{equation}

The elements of the Pfaff sequence for this Action become,
\begin{align}
A  & =xdy+dz.,\\
dA  & =dx\symbol{94}dy,\\
A\symbol{94}dA  & =dx\symbol{94}dy\symbol{94}dz,\\
dA\symbol{94}dA  & =0.
\end{align}
Note that for this example the coefficient of the 3-form of Topological
Torsion is not zero, and depends upon the Enstrophy (square of the Vorticity)
of the fluid flow.

\subsection{The Vector Processes}

Relative to the position vector $\mathbf{R}=[x,y,z]$ of ordered topological
coordinates $\{x,y,z\},$ consider the 3 abstract, linearly independent,
orthogonal (supposedly) vector\ direction fields:%
\begin{align}
\mathbf{V}_{x}  & =\left\vert
\begin{array}
[c]{c}%
1\\
0\\
0
\end{array}
\right\rangle ,\\
\mathbf{V}_{y}  & =\left\vert
\begin{array}
[c]{c}%
0\\
1\\
0
\end{array}
\right\rangle ,\\
\mathbf{E}  & =\left\vert
\begin{array}
[c]{c}%
0\\
0\\
1
\end{array}
\right\rangle .
\end{align}
These direction fields can be used to define a class of (real)\ Vector
processes, but these real vectors do not exhibit the complex Spinor class of
eigendirection fields for the 2-form, $dA$. \ The Spinor eigendirection fields
are missing from this basis frame. \ The important fact is that thermodynamic
processes defined in terms of a real basis frame (and its connection) are
incomplete, as such processes ignore the complex spinor direction fields.

For each of the real direction fields, deform the (assumed) process by an
arbitrary function, $\rho.$ \ Then construct the terms that make up the First
Law of topological thermodynamics. \ First construct the contractions to form
the internal energy for each process,%
\begin{align}
U_{\mathbf{V}_{x}}  & =i(\rho\mathbf{V}_{x})A=0,\text{ \ \ \ \ \ \ \ }%
dU_{\mathbf{V}_{x}}=0,\\
\ U_{\mathbf{V}_{y}}  & =i(\rho\mathbf{V}_{y})A=\rho x,\text{ \ }%
dU_{\mathbf{V}_{y}}=d(\rho x),\\
U_{\mathbf{E}}  & =i(\rho\mathbf{E})A=\rho,\text{ \ \ \ \ \ \ \ \ \ }%
dU_{\mathbf{E}}=d\rho.
\end{align}
The \textit{extremal} vector $\mathbf{E}$ is the unique eigenvector with
eigenvalue zero relative to the maximal rank antisymmetric matrix generated by
the 2-form, $dA$. \ The \textit{associated} vector $\mathbf{V}_{x}$ (relative
to the 1-form of Action, $A$, is orthogonal to the\ $y,z$\ plane. \ Recall
that any associated vector represents a local adiabatic process, as the Heat
flow is transverse to the process. \ The linearly independent thermodynamic
Work 1-forms for evolution in the direction of the 3 basis vectors are
determined to be,%
\begin{align}
W_{\mathbf{V}_{x}}  & =i(\rho\mathbf{V}_{x})dA=+\rho dy,\\
\ W_{\mathbf{V}_{y}}  & =i(\rho\mathbf{V}_{y})dA=-\rho dx,\\
W_{\mathbf{E}}  & =i(\rho\mathbf{E})dA=0.
\end{align}
From Cartan's Magic Formula representing the First Law as a description of
topological evolution,
\begin{equation}
L_{(\mathbf{V})}A=i(\rho\mathbf{V})dA+d(i(\rho\mathbf{V})A)\equiv Q,
\end{equation}
it becomes apparent that,%
\begin{align}
Q_{\mathbf{V}_{x}}  & =-\rho dy,\;\;\;\;\ \ \ \;\ \ dQ_{\mathbf{V}_{x}}%
=-d\rho\symbol{94}dy,\\
Q_{\mathbf{V}_{y}}  & =+xd\rho,\;\;\;\;\;\ \;dQ_{\mathbf{V}_{y}}%
=-d\rho\symbol{94}dx,\\
Q_{\mathbf{E}}  & =d\rho\;\;\;\;\;\ \ \ \ \ \ \ \ \ \ \ \ \ dQ_{\mathbf{E}}=0,
\end{align}
All processes in the extremal direction satisfy the conditions that
$Q_{\mathbf{E}}\symbol{94}dQ_{\mathbf{E}}=0.$ \ Hence, all extremal processes
are reversible. \ It is also true that evolutionary processes in the direction
of the other basis vectors, separately, are reversible, as the 3-form
$Q\symbol{94}dQ$ vanishes for $\mathbf{V}_{x},\ \mathbf{V}_{y},$ or
$\mathbf{E}$. \ Hence all such \textit{piecewise} continuous,
\textit{transitive}, processes are thermodynamically reversible. \ 

Note further that the "rotation" induced by the antisymmetric matrix $\left[
dA\right]  $\ acting on $\mathbf{V}_{x}$ yields $\mathbf{V}_{y}$ and the 4th
power of the matrix yields the identity rotation,%

\begin{align}
\left[  dA\right]  \circ\left\vert \mathbf{V}_{x}\right\rangle  & =\left\vert
\mathbf{V}_{y}\right\rangle ,\\
\left[  dA\right]  ^{2}\circ\left\vert \mathbf{V}_{x}\right\rangle  &
=-\left\vert \mathbf{V}_{x}\right\rangle ,\\
\left[  dA\right]  ^{4}\circ\left\vert \mathbf{V}_{x}\right\rangle  &
=+\left\vert \mathbf{V}_{x}\right\rangle .
\end{align}
This concept is a signature of Spinor phenomena.

\subsection{The Spinor Processes}

Now consider processes defined in terms of the Spinors. \ The eigendirection
fields of the antisymmetric matrix representation of $F=dA,$%

\begin{equation}
\left[  F\right]  =\left[
\begin{array}
[c]{ccc}%
0 & 1 & 0\\
-1 & 0 & 0\\
0 & 0 & 0
\end{array}
\right]  ,
\end{equation}
are given by the equations:
\begin{align}
\text{EigenSpinor1 \ }\left\vert Sp1\right\rangle  & =\left\vert
\begin{array}
[c]{c}%
1\\
\sqrt{\text{-1}}\\
0
\end{array}
\right\rangle ~\ \ \text{Eigenvalue }\text{= +}\sqrt{\text{-1}},\\
\text{EigenSpinor2 }\left\vert Sp2\right\rangle  & =\left\vert
\begin{array}
[c]{c}%
1\\
-\sqrt{\text{-1}}\\
0
\end{array}
\right\rangle ~\ \text{Eigenvalue }\text{= -}\sqrt{\text{-1}}\\
\text{EigenVector1}\left\vert E\right\rangle  & =\left\vert
\begin{array}
[c]{c}%
0\\
0\\
1
\end{array}
\right\rangle ~\ \ \ \ \ \ \text{Eigenvalue }\text{= 0}%
\end{align}

Now consider the processes defined by $\rho$ times the Spinor eigendirection
fields.\ Compute the change in internal energy, $dU$, the Work, $W$\ and the
Heat, $Q$, for each Spinor eigendirection field:\
\begin{align}
U_{\rho\mathbf{Sp}_{\mathbf{1}}}  & =i(\rho\mathbf{Sp}_{\mathbf{1}}%
)A=\sqrt{-1}\rho x\ \ \ \ \ \ \ \ \ \ \ d(U_{\rho\mathbf{Sp}_{\mathbf{1}}%
})=\sqrt{-1}d(\rho x),\\
U_{\rho\mathbf{Sp}_{\mathbf{2}}}  & =i(\rho\mathbf{Sp}_{\mathbf{2}}%
)A=-\sqrt{-1}\rho x\ \ \ \ \ \ \ \ d(U_{\rho\mathbf{Sp}_{\mathbf{2}}}%
)=-\sqrt{-1}d(\rho x),\\
U_{\rho\mathbf{E}}  & =i(\rho\mathbf{E})A=\rho
,\ \ \ \ \ \ \ \ \ \ \ \ \ \ \ \ \ \ \ \ \ \ d(U_{\rho\mathbf{E}})=d\rho,.
\end{align}

\begin{align}
\ W_{\rho\mathbf{Sp}_{\mathbf{1}}}  & =i(\rho\mathbf{Sp}_{\mathbf{1}}%
)dA=\rho(dy-\sqrt{\text{-1}}dx),\\
W_{\rho\mathbf{Sp}_{\mathbf{2}}}  & =i(\rho\mathbf{Sp}_{\mathbf{2}}%
)dA=+\rho(dy+\sqrt{\text{-1}}dx)\\
W_{\rho\mathbf{V}_{\mathbf{1}}}  & =i(\rho\mathbf{V}_{\mathbf{1}})dA=0,.
\end{align}

\begin{align}
\ Q_{\rho\mathbf{Sp}_{\mathbf{1}}}  & =L_{i(\rho\mathbf{Sp}_{\mathbf{1}}%
)}A=\rho(dy-\sqrt{\text{-1}}dx)+\sqrt{\text{-1}}d(\rho x),\\
Q_{\rho\mathbf{Sp}_{\mathbf{2}}}  & =L_{i(\rho\mathbf{Sp}_{\mathbf{2}})}%
A=\rho(dy+\sqrt{\text{-1}}dx)-\sqrt{\text{-1}}d(\rho x),\\
Q_{\rho\mathbf{V}_{\mathbf{1}}}  & =L_{i(\rho\mathbf{V}_{\mathbf{1}})}A=d\rho.
\end{align}

\subsection{Irreversible Spinor processes}

Next compute the 3-forms of $Q\symbol{94}dQ$ for each direction field,
including the spinors:%

\begin{align}
Q_{\rho\mathbf{V}_{\mathbf{1}}}\symbol{94}dQ_{\rho\mathbf{V}_{\mathbf{1}}}  &
=0,\\
\ Q_{\rho\mathbf{Sp}_{\mathbf{1}}}\symbol{94}dQ_{\rho\mathbf{Sp}_{\mathbf{1}%
}}  & =-\sqrt{\text{-1}}\rho d\rho\symbol{94}dx\symbol{94}dy,\\
Q_{\rho\mathbf{Sp}_{\mathbf{2}}}\symbol{94}dQ_{\rho\mathbf{Sp}_{\mathbf{2}}}
& =+\sqrt{\text{-1}}\rho d\rho\symbol{94}dx\symbol{94}dy.
\end{align}
It is apparent that evolution in the direction of the Spinor fields can be
irreversible in a thermodynamic sense, if $d\rho\symbol{94}dx\symbol{94}dy$ is
not zero. \ This is not true for the "piecewise linear"\ combinations of the
complex Spinors that produce the real vectors, $\mathbf{V}$ and $\mathbf{V}%
_{\bot}.$

Evolution in the direction of "smooth" combinations of the base vectors may
not satisfy the reversibility conditions, $Q\symbol{94}dQ=0$, when the
combination involves a fixed point in the $x,y$ plane. \ For example, it is
possible to consider smooth rotations (polarization chirality) in the $x,y$
plane: \
\begin{align}
V_{\text{rotation right}}  & =\mathbf{V}_{\bot}+\sqrt{\text{-1}}%
\mathbf{V=~}Sp1,\\
Q\symbol{94}dQ  & =-\sqrt{\text{-1}}\rho d\rho\symbol{94}dx\symbol{94}dy.
\end{align}%
\begin{align}
V_{\text{rotation left}}  & =\mathbf{V}-\sqrt{\text{-1}}\mathbf{V}_{\bot
}\mathbf{=~}Sp2,\\
Q\symbol{94}dQ  & =+\sqrt{\text{-1}}\rho d\rho\symbol{94}dx\symbol{94}dy.
\end{align}
The non-zero value of $Q\symbol{94}dQ$ for the continuous rotations are
related to the non-zero Godbillon-Vey class \cite{Pittie}. \ \ A key feature
of the rotations is that they have a fixed point in the plane; \ the motions
are not transitive. \ If the physical system admits an equation of state of
the form, $\theta=\theta(x,y,\rho)=0,$ then the rotation or expansion
processes are not irreversible. \ 

Note that the (supposedly) Vector processes of the preceding subsection are
combinations of the Spinor processes,
\begin{align}
\mathbf{V}_{x}  & =(a\cdot Sp1+b\cdot Sp2)/2\\
\mathbf{V}_{y}  & =-\sqrt{\text{-1}}(a\cdot Sp1-b\cdot Sp2)/2.
\end{align}
Almost always, a process defined in terms a linear combinations of the Spinor
direction fields will generate a Heat 1-form, $Q$, that does not satisfy the
Frobenius integrability theorem, and therefore all such processes are
thermodynamically irreversible: $Q\symbol{94}dQ\neq0$. \ \ However, with the
requirement that $a^{2}$ is precisely the same as $b^{2}$, then either
piecewise linear process is reversible, for $Q\symbol{94}dQ=0$. \ 

If the coefficients, and therefore the Spinor contributions, have slight
fluctuations, the cancellation of the complex terms is not precise. \ Then
either of the (now approximately) piecewise continuous process will NOT\ be
reversible due to Spinor fluctuations. \ \ 

\begin{remark}
\textit{The facts that piecewise (sequential) C1 transitive evolution along a
set of direction fields in odd (3) dimensions can be thermodynamically
reversible, }$Q\symbol{94}dQ=0$\textit{, while (smooth) C2 evolution processes
composed from complex Spinors can be thermodynamically irreversible,
\ }$Q\symbol{94}dQ\neq0$\textit{, is a remarkable result which appears to have
a relationship to Nash's theorem on C1 embedding. \ Physically, the results
are related to tangential discontinuities such as hydrodynamic wakes.}
\end{remark}

For systems of Pfaff dimension 4, all of the eigendirection fields are
Spinors. \ The Spinors occur as two conjugate pairs. \ If the conjugate
variables are taken to be x,y and z,t then the z,t spinor pair can be
interpreted in terms of a chirality of expansion or contraction, where the x,y
pair can be interpreted as a chirality of polarization. \ In this sense it may
be said that thermodynamic time irreversibility is an artifact of dimension 4. \ 

It is remarkable that a rotation and an expansion can be combined (eliminating
the fixed point) to produce a thermodynamically \textit{reversible} process.

Ian Stewart points out that there are three types of manifold structure:
piecewise linear, smooth, topological. \ Theorems on piecewise-linear
manifolds may not be true on smooth manifolds. \ The work above seems to
describe such an effect. \ Piecewise continuous processes are reversible,
where smooth continuous processes are not (see page 106, \cite{Stewart})!

\newpage

\section{Epilogue: \ Topological Fluctuations and Spinors}

This Section 5 goes beyond the original objective of demonstrating that the
Navier-Stokes equations, based upon continuous topological evolution, can
describe the irreversible decay of turbulence, but not its creation.
\ However, the key features of process irreversibility and turbulence are
entwined with the concept of Topological Torsion and Spinors. \ Hence this
epilogue calls attention to the fact that the Cartan topological methods
permit the analysis of Spinor entanglement, as well as the analysis of
fluctuations about kinematic perfection. \ This research area is in its
infancy, and extends the thermodynamic approach to the realm of fiber bundles.
\ A few of the introductory ideas are presented below.

\begin{remark}
These concepts go beyond the scope of this essay which has the objective of
presenting the important topological ideas in a manner palatable (if not
recognizable) to the engineering community of hydrodynamics.
\end{remark}

\subsection{The Cartan-Hilbert Action 1-form}

To start, consider those physical systems that can be described by a function
\textsf{L}$(\mathbf{q},\mathbf{v,}t)$ and a 1-form of Action given by
Cartan-Hilbert format,
\begin{equation}
A=\mathsf{L}(\mathbf{q}^{k},\mathbf{v}^{k}\mathbf{,}t)dt+\mathbf{p}%
_{k}\mathbf{\cdot}(d\mathbf{q}^{k}-\mathbf{v}^{k}dt).\label{Cart-Hilb}%
\end{equation}
The classic Lagrange function, \ \textsf{L}$(\mathbf{q}^{k},\mathbf{v}%
^{k}\mathbf{,}t)dt$, is extended to include fluctuations in the kinematic
variables, $(d\mathbf{q}^{k}-\mathbf{v}^{k}dt)\neq0$. \ It is no longer
assumed that the equation of Kinematic Perfection is satisfied. \ Fluctuations
of the topological constraint of Kinematic Perfection are permitted;
\begin{equation}
\text{\textbf{Topological Fluctuations in position:\ \ }}\Delta\mathbf{q}%
=(d\mathbf{q}^{k}-\mathbf{v}^{k}dt)\neq0.
\end{equation}
As the fluctuations are 1-forms, it is some interest to compute their Pfaff
Topological Dimension. \ The first step in the construction of the Pfaff
Sequence is to compute the exterior differential of the fluctuation 1-form:%

\begin{align}
\text{\textbf{Fluctuation 2-form}}\mathbf{:\ \ }  & \text{{}}d(\Delta
\mathbf{q)}=-(d\mathbf{v}^{k}-\mathbf{a}^{k}dt)\symbol{94}dt\\
& =-\Delta\mathbf{v}\symbol{94}dt,\\
\text{\textbf{Topological Fluctuations in velocity}}  & \mathbf{:}\text{{}%
}\Delta\mathbf{v=}(d\mathbf{v}^{k}-\mathbf{a}^{k}dt)\neq0.
\end{align}
It is apparent that the Pfaff Topological Dimension of the fluctuations is at
most 3, as $\Delta\mathbf{q\symbol{94}}\Delta\mathbf{v}\symbol{94}dt\neq0$,
and has a Heisenberg component,

When dealing with fluctuations in this prologue, the geometric dimension of
independent base variables will not be constrained to the 4 independent base
variables of the Thermodynamic model. \ At first glance it appears that the
domain of definition is a (3n+1)-dimensional variety of independent base
variables, $\{\mathbf{p}_{k},\mathbf{q}^{k},\mathbf{v}^{k}\mathbf{,}t\}$. \ Do
not make the assumption that the $\mathbf{p}_{k}$ are constrained to be
canonically defined. \ Instead, consider $\mathbf{p}_{k}$ to be a (set of)
Lagrange multiplier(s) to be determined later. \ Also, do not assume at this
stage that $\mathbf{v}$ is a kinematic velocity function, such that
$(d\mathbf{q}^{k}-\mathbf{v}^{k}dt)\Rightarrow0.$ \ The classical idea is to
assert that topological fluctuations in position are related to pressure, and
topological fluctuations in velocity are related to temperature.

For the given Action, construct the Pfaff Sequence (\ref{PS}) in order to
determine the Pfaff dimension or class \cite{PL1} of the Cartan-Hilbert 1-form
of Action. \ The top Pfaffian is defined as the non-zero p-form of largest
degree p in the sequence. \ The top Pfaffian for the Cartan-Hilbert Action is
given by the formula,
\begin{align}
& \text{\textbf{Top Pfaffian is 2n+2}}\nonumber\\
(dA)^{n+1}  & =(n+1)!\{\Sigma_{k=1}^{n}(\partial\mathsf{L}/\partial
v^{k}-p_{k})dv^{k}\}\symbol{94}\Omega_{2n+1},\\
\Omega_{2n+1}  & =dp_{1}\symbol{94}...dp_{n}\symbol{94}dq^{1}\symbol{94}%
..dq^{n}\symbol{94}dt.
\end{align}
The formula is a bit surprising in that it indicates that the Pfaff
Topological Dimension of the Cartan-Hilbert 1-form is 2n+2, and not the
geometrical dimension $3n+1$. \ For n = 3 "degrees of freedom", the top
Pfaffian indicates that the Pfaff Topological Dimension of the 2-form, $dA$ is
$2n+2=8$. \ The value $3n+1=10$ might be expected as the 1-form was defined
initially on a space of $3n+1$ "independent"\ base variables. \ The
implication is that there exists an irreducible number of independent
variables equal to $2n+2=8$ which completely characterize the differential
topology of the first order system described by the Cartan-Hilbert Action. It
follows that the exact 2-form, $dA,$ satisfies the equations%
\begin{equation}
(dA)^{n+1}\neq0,\,but\,\,\,\,\,A\symbol{94}(dA)^{n+1}=0.
\end{equation}

\begin{remark}
The idea that the 2-form, $dA$, is a symplectic generator of even maximal
rank, 2n+2, implies that ALL eigendirection fields of the 2-form, $F=dA$, are
complex isotropic Spinors, and all processes on such domains have Spinor components.
\end{remark}

The format of the top Pfaffian requires that the bracketed factor in the
expression above, $\{\Sigma_{k=1}^{n}(\partial$\textsf{L}$/\partial
v^{k}-p_{k})dv^{k}\},$ can be represented (to within a factor) by a perfect
differential, $dS.$%
\begin{equation}
dS=(n+1)!\{\Sigma_{k=1}^{n}(\partial\mathsf{L}/\partial v^{k}-p_{k})dv^{k}\}.
\end{equation}

The result is also true for any closed addition $\gamma\,\,$added to $A$;
e.g., the result is "gauge invariant". \ Addition of a closed 1-form does not
change the Pfaff dimension from even to odd. On the other hand the result is
not renormalizable, for multiplication of the Action 1-form by a function can
change the algebraic Pfaff dimension from even to odd.

On the 2n+2 domain, the components of (2n+1)-form $\,\,T=A\symbol{94}(dA)^{n}
$ generate what has been defined herein as the Topological Torsion vector, to
within a factor equal to the Torsion Current. \ The coefficients of the
(2n+1)-form are components of a contravariant vector density $\mathbf{T}^{m}$
defined as the Topological Torsion vector, the same concept as defined
previously on a 4D\ thermodynamic domain, but now extended to
(2n+2)-dimensions. \ This vector is orthogonal (transversal) to the 2n+2
components of the covector, $\mathbf{A}_{m}.$ In other words,%
\begin{equation}
A\symbol{94}T=A\symbol{94}(A\symbol{94}(dA)^{n})=0\Rightarrow i(\mathbf{T}%
)(A)=%
{\textstyle\sum}
\mathbf{T}^{m}\mathbf{A}_{m}=0.
\end{equation}
This result demonstrates that the extended Topological Torsion vector
represents an adiabatic process. \ This topological result does not depend
upon geometric ideas such as metric. \ It was demonstrated above that, on a
space of 4 independent variables, evolution in the direction of the
Topological Torsion vector is irreversible in a thermodynamic sense, subject
to the symplectic condition of non-zero divergence, $d(A\symbol{94}dA)\neq0. $
\ The same concept holds on dimension 2n+2.

The 2n+2 symplectic domain so constructed can not be compact without boundary
for it has a volume element which is exact. \ By Stokes theorem, if the
boundary is empty, then the surface integral is zero, which would require that
the volume element vanishes; but that is in contradiction to the assumption
that the volume element is finite. \ For the 2n+2 domain to be symplectic, the
top Pfaffian can never vanish. \ The domain is therefore orientable, but has
two components, of opposite orientation. \ Examination of the constraint that
the symplectic space be of dimension 2n+2 implies that the Lagrange
multipliers,\textbf{\ }$\mathbf{p}_{k}$, cannot be used to define momenta in
the classical "conjugate or canonical"\ manner.

\ Define the non-canonical components of the momentum, $\hslash k_{j},$ as,%
\begin{equation}
\text{\textbf{\ non-canonical momentum:} }\hslash k_{j}=(p_{j}-\partial
\mathsf{L}/\partial v^{j}),
\end{equation}
such that the top Pfaffian can be written as,{}%
\begin{align}
(dA)^{n+1}  & =(n+1)!\{\Sigma_{j=1}^{n}\hslash k_{j}dv^{j}\}\symbol{94}%
\Omega_{2n+1},\\
\Omega_{2n+1}  & =dp_{1}\symbol{94}...dp_{n}\symbol{94}dq^{1}\symbol{94}%
..dq^{n}\symbol{94}dt.
\end{align}

For the Cartan-Hilbert Action to be of Pfaff Topological Dimension 2n+2, the
factor $\{\Sigma_{j=1}^{n}\hslash k_{j}dv^{j}\}\neq0.$ \ It is important to
note, however, that as $(dA)^{n+1}$ is a volume element of geometric dimension
2n+2, the 1-form $\Sigma_{j=1}^{n}\hslash k_{j}dv^{j}$ is exact (to within a
factor, say $T(\mathbf{q}^{k},t,\ \mathbf{p}_{k}\mathbf{,}S_{\mathbf{v}})$);
hence,%
\begin{equation}
\Sigma_{j=1}^{n}\hslash k_{j}dv^{j}=TdS_{\mathbf{v}}.
\end{equation}
Tentatively, this 1-form, $dS_{\mathbf{v}},$ will be defined as the
Topological Entropy production relative to topological fluctuations of
momentum, kinematic differential position and velocity. \ \ If $\hslash k_{j}
$ is defined as the deviation about the canonical definition of momentum,
$\hslash k_{j}=\Delta p_{j},$ and noting the the expression for the top
Pfaffian can be written as $(n+1)!\{\Sigma_{j=1}^{n}\hslash k_{j}\Delta
v^{j}\}\symbol{94}\Omega_{2n+1}$, leads to an expression for the entropy
production rate in the suggestive "Heisenberg" format:%

\begin{equation}
TdS_{\mathbf{v}}=\Delta p_{j}\Delta v^{j}.
\end{equation}

\begin{quotation}
\bigskip
\end{quotation}

\end{document}